\documentclass[preprint,notoc]{JHEP3}

\usepackage{epsf}
\usepackage{psfrag}

\def\be{\begin{equation}}
\def\ee{\end{equation}}
\def\bea{\begin{eqnarray}}
\def\eea{\end{eqnarray}}

\def\eqn#1{eq.~(\ref{#1})}
\def\eqns#1#2{eqs.~(\ref{#1}) and~(\ref{#2})}
\def\fig#1{figure~{\ref{#1}}}
\def\Fig#1{Figure~{\ref{#1}}}

\def\e{\epsilon}

\def\oneloop{{\rm 1\! -\! loop}}

\def\Ord{{\cal O}}

\def\la{\langle}
\def\ra{\rangle}

\def\lr{\leftrightarrow}

\def\del{\partial}
\def\Psl{\not{\hbox{\kern-2.3pt $P$}}}
\def\psl{\not{\hbox{\kern-2.3pt $p$}}}
\def\Ksl{\not{\hbox{\kern-2.3pt $K$}}}
\def\ksl{\not{\hbox{\kern-2.3pt $k$}}}
\def\esl{\not{\hbox{\kern-2.3pt $\pol$}}}

\def\tr{\mathop{\rm tr}\nolimits}
\def\Tr{\mathop{\rm Tr}\nolimits}

\def\pol{\varepsilon}
\def\Neqfour{{\cal N}=4}
\def\Neqone{{\cal N}=1}

\def\spa#1.#2{\left\langle#1\,#2\right\rangle}
\def\spb#1.#2{\left[#1\,#2\right]}
\def\lor#1.#2{\left(#1\,#2\right)}
\def\sand#1.#2.#3{%
\left\langle\smash{#1}{\vphantom1}^{-}\right|{#2}%
\left|\smash{#3}{\vphantom1}^{-}\right\rangle}
\def\sandp#1.#2.#3{%
\left\langle\smash{#1}{\vphantom1}^{-}\right|{#2}%
\left|\smash{#3}{\vphantom1}^{+}\right\rangle}
\def\sandpp#1.#2.#3{%
\left\langle\smash{#1}{\vphantom1}^{+}\right|{#2}%
\left|\smash{#3}{\vphantom1}^{+}\right\rangle}
\def\sandpm#1.#2.#3{%
\left\langle\smash{#1}{\vphantom1}^{+}\right|{#2}%
\left|\smash{#3}{\vphantom1}^{-}\right\rangle}
\def\sandmp#1.#2.#3{%
\left\langle\smash{#1}{\vphantom1}^{-}\right|{#2}%
\left|\smash{#3}{\vphantom1}^{+}\right\rangle}
\def\sandmm#1.#2.#3{%
\left\langle\smash{#1}{\vphantom1}^{-}\right|{#2}%
\left|\smash{#3}{\vphantom1}^{-}\right\rangle}
\def\spab#1.#2.#3{\sandmm#1.#2.#3}
\def\spbb#1.#2.#3.#4{\sandpm#1.{#2#3}.#4}
\newbox\charbox
\newbox\slabox
\def\s#1{{      % Feynman slash
        \setbox\charbox=\hbox{$#1$}
        \setbox\slabox=\hbox{$/$}
        \dimen\charbox=\ht\slabox
        \advance\dimen\charbox by -\dp\slabox
        \advance\dimen\charbox by -\ht\charbox
        \advance\dimen\charbox by \dp\charbox
        \divide\dimen\charbox by 2
        \raise-\dimen\charbox\hbox to \wd\charbox{\hss/\hss}
        \llap{$#1$}
}}
\def\ksl{\s{k}}
\def\lsl{\s{l}}
\def\delsl{\s{\del}}
\def\Dsl{\s{D}}

\def\beqa{\begin{eqnarray}}
\def\eeqa{\end{eqnarray}}
\def\beq{\begin{equation}}
\def\eeq{\end{equation}}
\def\hf{{\textstyle{1\over2}}}
\def\ihf{{\textstyle{i\over2}}}
\def\sst{\scriptscriptstyle}

\def\vev#1{\langle{#1}\rangle}

\preprint{
  hep-th/0411092\\
  SLAC--PUB--10839\\
  IPPP/04/71\\
  DCPT/04/142\\
  November, 2004}

\title{MHV Rules for Higgs Plus Multi-Gluon Amplitudes}

\author{Lance J. Dixon,$^{(a,b)}$\thanks{%
Research supported by the US Department of
Energy under contract DE-AC02-76SF00515.}
    \ E. W. N. Glover$^{(b)}$ and Valentin V. Khoze$^{(b)}$\\
    $^{(a)}$ Stanford Linear Accelerator Center, Stanford University,
    Stanford, CA 94309, USA\\
        $^{(b)}$
        Institute of Particle Physics Phenomenology,
        Department of Physics,\\
        University of Durham, Durham, DH1 3LE, UK\\
        E-mail: \email{lance@slac.stanford.edu, e.w.n.glover@durham.ac.uk, 
        valya.khoze@durham.ac.uk}}

\abstract{
We use tree-level perturbation theory to show how non-supersymmetric
one-loop scattering amplitudes for a Higgs boson plus an arbitrary number of
partons can be constructed, in the limit of a heavy top quark, 
from a generalization of the scalar graph approach of Cachazo, Svr\v{c}ek 
and Witten.  The Higgs boson couples to gluons through a top quark loop 
which generates, for large $m_t$, a dimension-5 operator 
$H \tr G_{\mu\nu} G^{\mu\nu}$.   This effective interaction leads to
amplitudes which cannot be described by the standard MHV rules; for
example, amplitudes where all of the gluons have positive helicity.
We split the effective interaction into the sum of two terms, one holomorphic
(selfdual) and one anti-holomorphic (anti-selfdual). 
The holomorphic interactions give a new set of MHV vertices 
--- identical in form to those of pure gauge theory, except for 
momentum conservation  --- that can be combined with
pure gauge theory MHV vertices to produce a tower of amplitudes with more
than two negative helicities.   Similarly, the anti-holomorphic
interactions give anti-MHV vertices that can be combined with
pure gauge theory anti-MHV vertices  to produce a tower of amplitudes with
more than two positive helicities.  A Higgs boson amplitude is the sum
of one MHV-tower amplitude and one anti-MHV-tower amplitude.
We present all MHV-tower amplitudes with up to
four negative-helicity gluons and any number of positive-helicity 
gluons (NNMHV). 
These rules reproduce all of the available analytic 
formulae for Higgs $+$ $n$-gluon scattering ($n\leq5$) at tree level, 
in some cases yielding considerably shorter expressions. }

\keywords{QCD, Higgs boson, Supersymmetry and Duality, Hadron Colliders}

\dedicated{\centerline{Submitted to JHEP}}

\begin{document}

%%%%%%%%%%%%%%%%%%%%%%%%%%%%%%%%%%%%%%%%%%%%%%%%%%%

\section{Introduction}
\label{IntroSection}

Since the interpretation of $\Neqfour$ supersymmetric Yang-Mills theory
and QCD as a topological string propagating in twistor
space~\cite{Witten1} (at least at tree level), there has been a flurry of
activity.
Very notably, a new set of `MHV rules'
has been proposed for QCD~\cite{CSW1}, which take the place of ordinary
Feynman rules, and lump many Feynman vertices into single color-ordered
`MHV vertices'.  These MHV vertices are off-shell continuations of the
maximally helicity-violating (MHV) $n$-gluon scattering amplitudes 
of Parke and Taylor~\cite{ParkeTaylor}.  Written in terms of spinor inner 
products~\cite{SpinorHelicity}, they are composed entirely of the `holomorphic'
products $\spa{i}.{j}$ fashioned from right-handed (undotted) spinors, 
rather than their anti-holomorphic partners $\spb{i}.{j}$,
\be
 A_n(1^+,\ldots,p^-,\ldots,q^-,\ldots,n^+) =
{ \spa{p}.{q}^4 \over \spa1.2 \spa2.3 \cdots \spa{n-1,}.{n} \spa{n}.{1} }.
\label{MHV}
\ee
Here $p$ and $q$ are the only gluons with negative helicity.
The MHV vertices~(\ref{MHV}), with a suitable definition for 
$\spa{i}.{j}$ when the momenta $k_i$ or $k_j$ are off shell~\cite{CSW1},
are then connected with scalar-type propagators, which bear the helicity 
of the intermediate leg but no Lorentz indices.
In twistor space, where the anti-holomorphic 
spinors $\tilde\lambda_{i,\dot\alpha}$ are traded for their 
Fourier-transform variables 
$\mu_i^{\dot\alpha} = -i\partial/\partial\tilde\lambda_{i,\dot\alpha}$,
each MHV vertex is localized on a line.  The lines are connected
through the off-shell propagators.

The CSW approach has been extended to amplitudes with fermions~\cite{GK}.  
New tree-level gauge-theory results were obtained in this
approach for non-MHV amplitudes involving gluons~\cite{KosowerNMHV,BBK},
and gluons, fermions and scalars~\cite{GK,GGK,VVK}.

Although the topological string appears to mix gauge theory with conformal
supergravity at the loop level~\cite{BW}, the MHV rules nevertheless work
at the one-loop level in supersymmetric Yang-Mills theories (SYM).
Brandhuber, Spence and Travaglini~\cite{BST} used MHV rules to reproduce
the series of one-loop $n$-gluon MHV amplitudes in $\Neqfour$ SYM,
previously computed via unitarity cuts~\cite{BDDK1}.
Very recently, this approach was shown to work also for the same 
amplitudes in $\Neqone$ SYM~\cite{QuigleyRozali,BBST}.
On the other hand, the twistor-space structure of both sets of
amplitudes seemed to be more complex~\cite{CSW2} than the MHV picture would
imply.  This paradox was resolved by the notion of
a `holomorphic anomaly'~\cite{CSW3} due to singularities 
in the loop-momentum integration.  The anomaly for a unitarity cut
freezes the phase-space integration, making its evaluation 
simple~\cite{BBKR,Cachazo}.
In $\Neqfour$ SYM, the anomaly can be used to derive algebraic
equations for the coefficients of integral functions~\cite{Cachazo,BCF},
whose solutions are in agreement with a direct evaluation of the 
unitary cuts for 7-gluon amplitudes~\cite{BDDK7}.  
In $\Neqone$ SYM, differential instead of algebraic equations 
are obtained~\cite{BBDD}.

Much of this progress at one loop in massless supersymmetric theories is
related to the fact that such theories are
`cut-constructible'~\cite{BDDK2}.  That is, at one loop, intermediate
states can be assigned four-dimensional helicities~\cite{BDDK2}, even
though the loop-momentum integral must be regulated dimensionally, with
$D=4-2\e$.  The `error' incurred by neglecting the $(-2\e)$ components
of the momentum in numerators of the cuts can be confined to terms that
vanish as $\e \to 0$.

Application of MHV rules to loop amplitudes in {\it non-supersymmetric}
theories seems to be a different matter.  This situation is highlighted by
the properties of the $n$-gluon one-loop amplitudes for which all gluons 
(or all but one) have the same helicity, namely
${\cal A}_{n}^{\oneloop}(1^\pm,2^+,3^+,\ldots,n^+)$%
~\cite{BK1,PolyAmps,AllPlus,MahlonOneMinus}.
Such amplitudes vanish in the supersymmetric case, but are nonzero for
nonsupersymmetric combinations of massless gluons, fermions or 
scalars circulating in the loop.  They are finite as $\e\to0$, and
in this limit they become cut-free, rational functions of the kinematic 
invariants.  These amplitudes can still be computed from 
unitarity cuts in $D$ dimensions, by working to $\Ord(\e)$ or higher, but 
now the full $D$-dimensional set of intermediate states
enter~\cite{BDKAnnualReview,SelfDual}. Indeed, in a cut-based construction 
these amplitudes have support only when the loop momenta are {\it not} 
four-dimensional, but point into the $(-2\e)$ directions 
of $(D=4-2\e)$-dimensional space-time.  Assigning a four-dimensional helicity
to a state circulating around the loop seems unlikely to lead to a correct
answer in this case.  On the other hand, the quasi-local
nature of the $({\pm}{+}{+}\cdots{+})$ amplitudes suggests that some of them
might be considered fundamental vertices~\cite{CSW2}, like the tree-level
MHV vertices.  However, it has not yet been possible to find suitable
off-shell continuations, possibly because of the existence of
`anti-holomorphic' spinor products $\spb{i}.{j}$ in the numerators of the
amplitudes (as we shall review shortly).

A similar problem has plagued attempts to construct MHV
rules for gravity~\cite{GRRT}.  Tree-level gravity amplitudes
can be constructed from tree-level gauge-theory amplitudes using low-energy
limits of the Kawai-Lewellen-Tye relations in string theory~\cite{KLT}.
The gravity amplitudes are the sums of products of pairs of
gauge theory amplitudes, but there are additional factors of
$s_{ij} = (k_i+k_j)^2 = 2k_i\cdot k_j$ in the numerator.
Because $s_{ij} = \spa{i}.{j} \spb{j}.{i}$, anti-holomorphic spinor
products also creep in here.

In any case, it is of interest to extend the range of processes that
can be treated by MHV-type techniques. One set of processes
of much phenomenological interest is the scattering
of a single Higgs boson together with a number of quarks and
gluons. In fact, production of the Standard Model Higgs boson
at hadron colliders such as the Fermilab Tevatron and the CERN Large
Hadron Collider is dominated by gluon-gluon fusion, $gg\to H$, through
a one-loop diagram containing the top quark in the loop.
Precision electroweak data, interpreted within the Standard Model,
indicate that the Higgs is considerably lighter than $2m_t \approx 360$~GeV;
currently $m_H < 260$~GeV at 95\% confidence level~\cite{Renton:2004wd}.
In this case, next-to-leading-order QCD computations have
shown~\cite{NLOHiggs} that it is an excellent approximation to
integrate out the heavy top quark, summarizing its effects via the
dimension-five operator 
$H \tr G_{\mu\nu} G^{\mu\nu}$~\cite{HggOperator,HggOperator2}.
This operator can then be `dressed' by standard QCD vertices in order 
to generate Higgs plus multi-parton amplitudes.
In this paper, we will provide a set of MHV rules for such amplitudes.

The amplitudes for a Higgs boson plus four gluons ($Hgggg$) were first 
computed in the heavy top quark approximation by Dawson and 
Kauffman~\cite{DawsonKauffman}.
Kauffman, Desai and Risal extended these results to the other 
four-parton processes, $Hggq\bar{q}$ and $Hq\bar{q}Q\bar{Q}$~\cite{KDR}.
More recently, analytic formulae for a Higgs plus up to 5 partons were 
calculated by Del Duca, Frizzo and Maltoni~\cite{DFM}.  Amplitudes for 
these cases, and those with larger numbers of partons, are also computed 
numerically by the programs Alpha~\cite{Alpha} and MadGraph~\cite{MadGraph}.

An interesting subset of the Higgs plus $n$-gluon (color-ordered)
amplitudes are those for which all gluons have positive 
helicity,
\be
 A_n(H,1^+,2^+,\ldots,n^+) \propto
{ m_H^4 \over \spa1.2 \spa2.3 \cdots \spa{n-1,}.{n} \spa{n}.{1} } \,,
\label{HAllPlus}
\ee
where $m_H$ is the mass of the Higgs boson.\footnote{%
We shall prove this result for all $n$ in Appendix~\ref{VanishingSection}.
}
There is a strong similarity between the sequence~(\ref{HAllPlus})
and the (leading-color, color-ordered) pure QCD one-loop amplitudes 
for $n$ positive-helicity gluons~\cite{AllPlus,MahlonOneMinus},
\be
 A_{n;1}^{\oneloop}(1^+,2^+,\ldots,n^+)\propto
{ \sum_{1\leq i_1 < i_2 < i_3 < i_4 \leq n}
\spa{i_1}.{i_2} \spb{i_2}.{i_3} \spa{i_3}.{i_4} \spb{i_4}.{i_1}
\over \spa1.2 \spa2.3 \cdots \spa{n-1,}.{n} \spa{n}.{1} } \,.
\label{gAllPlus}
\ee 
Both sets of amplitudes are generated first at one loop (if we count
the top-quark loop in the Higgs case). They are both rational functions of
the kinematic variables; {\it i.e.} they contain no branch cuts.  (For the
Higgs case, as we work in the heavy top-quark limit, this statement is
rather trivial.) Their collinear and multi-particle factorization
properties are very similar, as reflected in the factors in their
denominators, $\spa1.2\spa2.3 \cdots \spa{n}.1$.  The numerator factors are
also quite similar, when momentum conservation is taken into account.  In the
Higgs case, the numerator factor is $m_H^4$; but this is also expressible
in terms of the $n$ gluon momenta as 
$ (\sum_{1\leq i < j \leq n} s_{ij})^2 
= (\sum_{1\leq i < j \leq n} \spa{i}.{j} \spb{j}.{i} )^2$.
Thus both the all-plus Higgs and one-loop QCD amplitudes are bi-linear
in the anti-holomorphic spinor products $\spb{i}.{j}$.

In the one-loop pure-gauge-theory case, a well-motivated but unsuccessful
attempt was made~\cite{CSW2} to generate the one-loop 5-point amplitude 
$A_{5;1}^{\oneloop}({-}{+}{+}{+}{+})$ from an
off-shell continuation of the one-loop 4-point amplitude
$A_4^{\oneloop}({+}{+}{+}{+})$, plus the off-shell tree-level 
MHV vertex $A_3({-}{-}{+})$.  In the Higgs case, we have made an 
analogous attempt to generate the Higgs plus three gluon
amplitude $A_3(H,{-}{+}{+})$ from an off-shell continuation of the
Higgs plus two gluon amplitude $A_2(H,{+}{+})$, plus the tree-level MHV vertex
$A_3({-}{-}{+})$.  Our attempt failed; it led to results which depended
on the choice of the `reference spinor' in the CSW construction, 
and thus could not be correct.

For the Higgs case, we shall resolve this problem in a different way, 
yet still using an MHV-type perturbation theory.  As we shall explain
in the next section, the crux of our method is to split the 
$H \tr G_{\mu\nu} G^{\mu\nu}$ operator into two terms.  
We will show that MHV rules can be applied to amplitudes
generated by one of the two terms. The amplitudes generated by the other
term can be obtained by applying `anti-MHV rules' (or deduced from the 
first set of amplitudes using parity).   The desired $Hgg\ldots g$ 
amplitude is the sum of one amplitude of each type.
The split of $H \tr G^2$ into two operators can be motivated either 
by supersymmetry, or by selfduality.
The general structure we find can be extended to Higgs amplitudes 
containing quarks~\cite{WIP}, 
although we shall not do so explicitly in this paper.

We hope that the MHV structure we have uncovered for the Higgs plus
multi-parton amplitudes will prove useful for understanding how to
apply twistor-MHV methods to one-loop amplitudes in pure QCD, 
and to tree-level amplitudes in gravity.
In the meantime, it allows us to obtain relatively compact expressions
for scattering amplitudes of phenomenological interest, such
as $gg\to Hggg$.
For example, the computation of the cross section for Higgs production 
via gluon-gluon fusion at nonzero transverse momentum 
(which may alleviate the QCD background in the $H\to \gamma\gamma$ decay 
mode~\cite{Hgammagamma,HiggsPT}) 
at next-to-leading order (NLO) in $\alpha_s$ requires 
amplitudes like $gg \to Hgg$; at next-to-next-to-leading order, it
requires $gg \to Hggg$.   The process 
$gg \to Hgg$~\cite{H2j,H2j2} also appears at leading order
(and so $gg \to Hggg$ will be needed at NLO) 
as a background to production of a Higgs boson via weak boson fusion, 
$q\bar{Q} \to q'\bar{Q}' W^+ W^- \to q'\bar{Q}' H$~\cite{WBF}. 
In both cases there are two additional jets, which are used to tag the 
weak boson fusion production process.  
The weak boson fusion process is now known at NLO~\cite{NLOWBFFOZ,NLOWBFBC},
but the gluon-fusion background is currently only available at leading
order, with large uncertainties~\cite{NLOWBFBC}.
The background is particularly important if one wants to use 
azimuthal correlations between the tagging jets to learn about the 
couplings of the Higgs boson to $W$ pairs~\cite{H2j2,Odagiri}.   
The compact formulae for tree-level Higgs amplitudes
should speed up their computation in numerical programs for higher-order
cross sections, where they may need to be evaluated very often.

This paper is organized as follows:
In Section~\ref{ModelSection}, we explain the method in more detail.  
The new MHV rules are summarized in Section~\ref{MHVRulesSection}. 
Using these rules, we obtain results for all Next-to-MHV and Next-to-Next-to-MHV 
amplitudes (with up to four negative-helicity gluons and any number 
of positive-helicity gluons).  We show that these results can be used to 
reproduce all of the available analytic formulae for Higgs $+$ $n$-gluon 
scattering ($n\leq5$) at tree level.  We also reproduce the all-$n$
formula~(\ref{HAllPlus}), for the special case where all gluons have the same
helicity.  It should be straightforward to 
apply our method to $n > 5$ partons and obtain new analytic results for 
amplitudes for a Higgs boson plus six partons~\cite{WIP}.

In Section~\ref{AnotherModel} we consider another example of an effective
theory describing gluonic interactions, where the tree-level all-plus
helicity amplitude does not vanish, but can again be reconstructed using
new MHV rules of the type presented in Sections~\ref{ModelSection} 
and \ref{MHVRulesSection}.  Our findings are summarized in the Conclusions.
There are two Appendices.  Appendix~\ref{Nots} summarizes our conventions
for color, spinors, helicity and selfduality.
Appendix~\ref{VanishingSection} describes technical details
necessary to show the triviality of certain classes of amplitudes.
It also contains the recursive construction of the infinite 
sequence of non-vanishing identical-helicity amplitudes given
in \eqn{HAllPlus}.

%%%%%%%%%%%%%%%%%%%%%%%%%%%%%%%%%%%

\section{The model}
\label{ModelSection}

In the Standard Model the Higgs boson couples to gluons through 
a fermion loop. The dominant contribution is from the top quark.  
For large $m_t$, the top quark can be integrated out, leading 
to the effective interaction~\cite{HggOperator,HggOperator2},
 \be
 {\cal L}_{\sst H}^{\rm int} = 
  {C\over2}\, H \tr G_{\mu\nu} G^{\mu\nu}  \ .
 \label{HGGeff}
 \ee
In the Standard Model, and to leading order in $\alpha_s$, 
the strength of the interaction is given by $C = \alpha_s/(6\pi v)$, 
with $v = 246$~GeV.  

The MHV or twistor-space structure of the Higgs-plus-gluons amplitudes is best
elucidated by dividing the Higgs coupling to gluons, \eqn{HGGeff}, into
two terms, containing purely selfdual (SD) and purely anti-selfdual (ASD)
gluon field strengths,
\be
G_{\sst SD}^{\mu\nu} = \hf(G^{\mu\nu}+{}^*G^{\mu\nu}) \ , \quad
G_{\sst ASD}^{\mu\nu} = \hf(G^{\mu\nu}-{}^*G^{\mu\nu}) \ , \quad
{}^*G^{\mu\nu} \equiv \ihf \epsilon^{\mu\nu\rho\sigma} G_{\rho\sigma} \ .
\ee
This division can be accomplished by considering $H$ to be the real 
part of a complex field $\phi = {1\over2}( H + i A )$, so that
\bea
 {\cal L}^{\rm int}_{H,A} &=& 
{C\over2} \Bigl[ H \tr G_{\mu\nu} G^{\mu\nu} 
             + i A \tr G_{\mu\nu}\, {}^*G^{\mu\nu} \Bigr]
 \label{effinta}\\
&=&
C \Bigl[ \phi \tr G_{{\sst SD}\,\mu\nu} G_{\sst SD}^{\mu\nu}
 + \phi^\dagger \tr G_{{\sst ASD}\,\mu\nu} G_{\sst ASD}^{\mu\nu} \Bigr]
 \ .
 \label{effintb}
\eea
The key idea is that, due to selfduality, the amplitudes for 
$\phi$ plus $n$ gluons, and those for $\phi^\dagger$ plus $n$ gluons, 
separately have a simpler structure than the amplitudes for 
$H$ plus $n$ gluons.  But because $H = \phi + \phi^\dagger$, 
the Higgs amplitudes can be recovered 
as the sum of the $\phi$ and $\phi^\dagger$ amplitudes.

As another motivation for the split~(\ref{effintb}), note that
this interaction can be embedded into an $\Neqone$ supersymmetric 
effective Lagrangian,
\be
{\cal L}^{\rm int}_{\rm SUSY} = 
- C \int d^2\theta\ \Phi \tr W^\alpha W_\alpha \, 
- C \, \int d^2\bar\theta\ \Phi^{\dagger} 
 \tr \overline{W}_{\dot\alpha}\overline{W}^{\dot\alpha} \,.
\label{susyembd}
\ee
Here $G_{\sst SD}^{\mu\nu}$ is the bosonic component of the chiral
superfield $W_\alpha$, and $\phi$ is the lowest component of the chiral
superfield $\Phi$.  In Appendix~\ref{Nots} we identify the
following helicity assignments:
\bea
&& W_\alpha = \{ g^{-},\lambda^{-} \} \ , \qquad 
\Phi = \{ \phi,\psi^{-} \} \ , \\
&& \overline{W}^{\dot\alpha} = \{g^{+}, \lambda^{+} \} \ , \qquad
\Phi^{\dagger} = \{\phi^{\dagger},\psi^{+}\} \ ,
\label{Helass}
\eea
where $g^{\pm}$ correspond to gluons with $h=\pm 1$ helicities,
$\lambda^{\pm}$ are gluinos with $h=\pm 1/2$,
$\phi$ and $\phi^{\dagger}$ are complex scalar
fields, and $\psi^{\pm}$ are their fermionic superpartners.
In Appendix~\ref{VanishingSection} we give the full supersymmetric
effective Lagrangian.  (This Lagrangian can be generated from a 
renormalizable, supersymmetric microscopic theory containing a 
massive top quark/squark chiral multiplet $T$, coupled to $\Phi$ by a Yukawa 
coupling $\int d^2\theta \Phi T \overline{T}$.  Integrating out $T$
produces the interaction~(\ref{susyembd}) with a coefficient proportional
to the chiral multiplet's contribution to the SYM beta function.)

As in the case of QCD, the fermionic superpartners of the boson 
$\phi$ and of the gluons will never enter tree-level processes for 
$\phi$ plus $n$ gluons.  Thus these bosonic amplitudes must obey
supersymmetry Ward identities (SWI)~\cite{SWI}, as discussed in 
Appendix~\ref{VanishingSection}, which help to control their structure.
Although we do not explicitly describe Higgs amplitudes involving 
quarks in this paper, we note that massless quarks can always be 
added to \eqn{susyembd} in a supersymmetric fashion, as members of 
additional chiral multiplets with no superpotential.  
Then the squarks, as well as the other superpartners, do not enter 
the amplitudes for a Higgs boson plus multiple quarks and gluons. 

We use a standard, trace-based color decomposition for the tree-level 
amplitudes for $n$ gluons, plus the single colorless field,
$H$, $\phi$ or $\phi^\dagger$.  The full amplitudes
can be assembled from color-ordered partial amplitudes $A_n$, as 
described in Appendix~\ref{Nots}.
Because all amplitudes calculated in this paper are proportional to one 
power of the constant $C$ appearing in \eqn{HGGeff}, we also remove $C$ 
and powers of the gauge coupling $g$ from the partial amplitudes $A_n$,
via the color decomposition formula~(\ref{TreeColorDecomposition}).

As proven in Appendix~\ref{VanishingSection}, a host of helicity 
amplitudes involving $\phi$ vanish, namely $\phi g^\pm g^+ g^+ \ldots g^+$. 
One version of the proof invokes the supersymmetry Ward 
identities~\cite{SWI}, but is valid only for a massless Higgs,
$m_H=0$.  The other version uses recursive techniques and the  
Berends-Giele currents~\cite{BerendsGiele}, and is valid for any $m_H$.
Thus we know that
\be
  A_n(\phi,1^\pm,2^+,3^+,\ldots,n^+) = 0 \,,
 \label{phimpvanish}
\ee
for all $n$.  

The MHV amplitudes, with precisely two negative helicities, 
$\phi g^- g^+ \ldots g^+ g^- g^+ \ldots  g^+$, are the 
first non-vanishing $\phi$ amplitudes.
General factorization properties now imply that they have to be extremely
simple.  They can have no multi-particle poles, because the residue
of such a pole would have to be the product of two $\phi$ amplitudes
with a total of three external negative-helicity gluons (one is 
assigned to the intermediate state).  At least one of the two product 
amplitudes must vanish according to \eqn{phimpvanish}.
Similarly, only factors of $\spa{i}.{j}$, not $\spb{i}.{j}$, are allowed 
in the denominator.  This property follows from the collinear limit
as $k_i$ becomes parallel to $k_j$; $\spb{i}.{j}$ factors are
associated with collinear factorization onto vanishing $(n-1)$-point
amplitudes of the form~(\ref{phimpvanish}).  
These same conditions are obeyed by MHV amplitudes in pure QCD;
indeed the arguments are identical.

Furthermore, the first few known $\phi$ amplitudes have precisely the same form
as the QCD case --- except for the implicit momentum carried out of the
process by the Higgs boson.  (This momentum makes the Higgs case well-defined
on-shell for fewer legs than the pure QCD case.)  We have,
\bea
  A_2(\phi,1^-,2^-) &=& { {\spa1.2}^4 \over \spa1.2 \spa2.1 } = - {\spa1.2}^2,
 \label{Hmm} \\
  A_3(\phi,1^-,2^-,3^+) &=& { {\spa1.2}^4 \over \spa1.2 \spa2.3 \spa3.1 }
                    = { {\spa1.2}^3 \over \spa2.3 \spa3.1 } \,,
 \label{Hmmp} \\
 A_4(\phi,1^-,2^-,3^+,4^+) &=& { {\spa1.2}^4 \over \spa1.2\spa2.3\spa3.4\spa4.1 }
                   \,.
 \label{Hmmpp}
\eea
This leads to the obvious assertion for {\it all} `$\phi$-MHV' amplitudes,
\be
A_n(\phi,1^+,2^+,\ldots, p^-, \ldots, q^-, \ldots ,n^+) =
 { {\spa{p}.{q}}^4 \over \spa1.2 \spa2.3 \cdots \spa{n-1,}.{n} \spa{n}.1 } \,,
\label{assrtn}
\ee
where only legs $p$ and $q$ have negative helicity.
Besides the correct collinear and multi-particle factorization 
behavior, these amplitudes also correctly reduce to pure QCD 
amplitudes as the $\phi$ momentum approaches zero.
It should be possible to prove~\eqn{assrtn} recursively,
along the lines of the proof in the QCD case~\cite{BerendsGiele},
perhaps using the additional light-cone recursive currents from
ref.~\cite{LightConeRecursive}.

Since the MHV amplitudes~(\ref{assrtn}) have an identical form to the 
corresponding amplitudes of pure Yang-Mills theory, \eqn{MHV}, 
we propose that their off-shell continuation is also identical to 
that proposed in the pure-glue context, in the context of a set of 
scalar-graph rules~\cite{CSW1}.
Everywhere the off-shell leg $i$ carrying momentum $K_i$ appears
in \eqn{assrtn}, we let the corresponding holomorphic spinor be
$\lambda_{i,\alpha} = (K_i)_{\alpha\dot\alpha} \xi^{\dot\alpha}$.
Here $\xi^{\dot\alpha}$ is an arbitrary reference spinor, chosen to be
the same for all MHV diagrams contributing to the amplitude.
(Because anti-holomorphic spinors $\lambda_{i,\dot\alpha}$ do not appear 
in~\eqn{assrtn}, we do not have to discuss their off-shell continuation.)

%%%%%%%%%%%%%%%%%%%%
%FIGURE
%
\FIGURE[th!]{
{\epsfxsize 4 truein \epsfbox{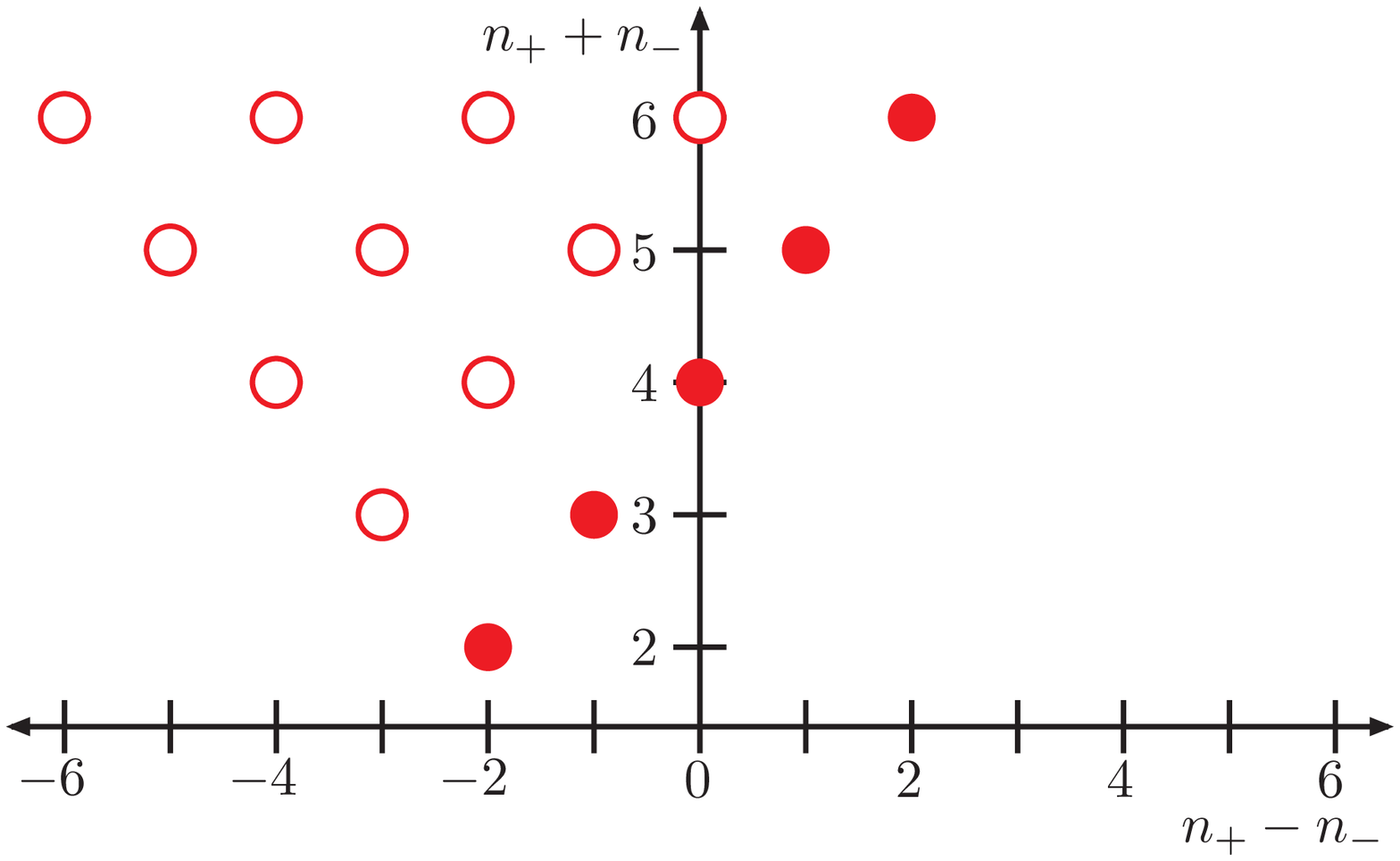}}
\caption{The MHV structure of $\phi$ plus multi-gluon amplitudes.
The number of positive (negative) helicity gluons is $n_+$ ($n_-$).
The vertical axis labels the total number of gluons, $n_+ + n_-$.
The horizontal axis labels the difference $n_+ - n_-$, a measure
of the amount of `helicity violation'.
Solid red dots represent fundamental `$\phi$-MHV' vertices.
Open red circles represent $\phi$ amplitudes which are composites,
built from the $\phi$-MHV vertices plus pure-gauge-theory MHV vertices.
}
\label{PhiTreeMapFigure}
}

\Fig{PhiTreeMapFigure} lays out the MHV structure of the $\phi$ plus
multi-gluon amplitudes.   All non-vanishing amplitudes are labelled with
circles.  The  fundamental $\phi$-MHV vertices, which coincide with the $\phi
g^-g^- g^+ \ldots g^+$ amplitudes~(\ref{assrtn}), are the basic building 
blocks and are labelled by red dots. The result of combining 
$\phi$-MHV vertices with pure-gauge-theory MHV vertices is to produce 
amplitudes with more than two negative helicities. These amplitudes 
are represented as red open circles.  Each MHV diagram contains 
exactly one $\phi$-MHV vertex; the rest are pure-gauge-theory MHV
vertices.  The vertices are combined with scalar propagators.
The MHV-drift is always to the left and upwards.  Collectively, these
amplitudes form the holomorphic (or MHV) tower of accessible amplitudes.

%%%%%%%%%%%%%%%%%%%%
%FIGURE
%
\FIGURE[th!]{
{\epsfxsize 4 truein \epsfbox{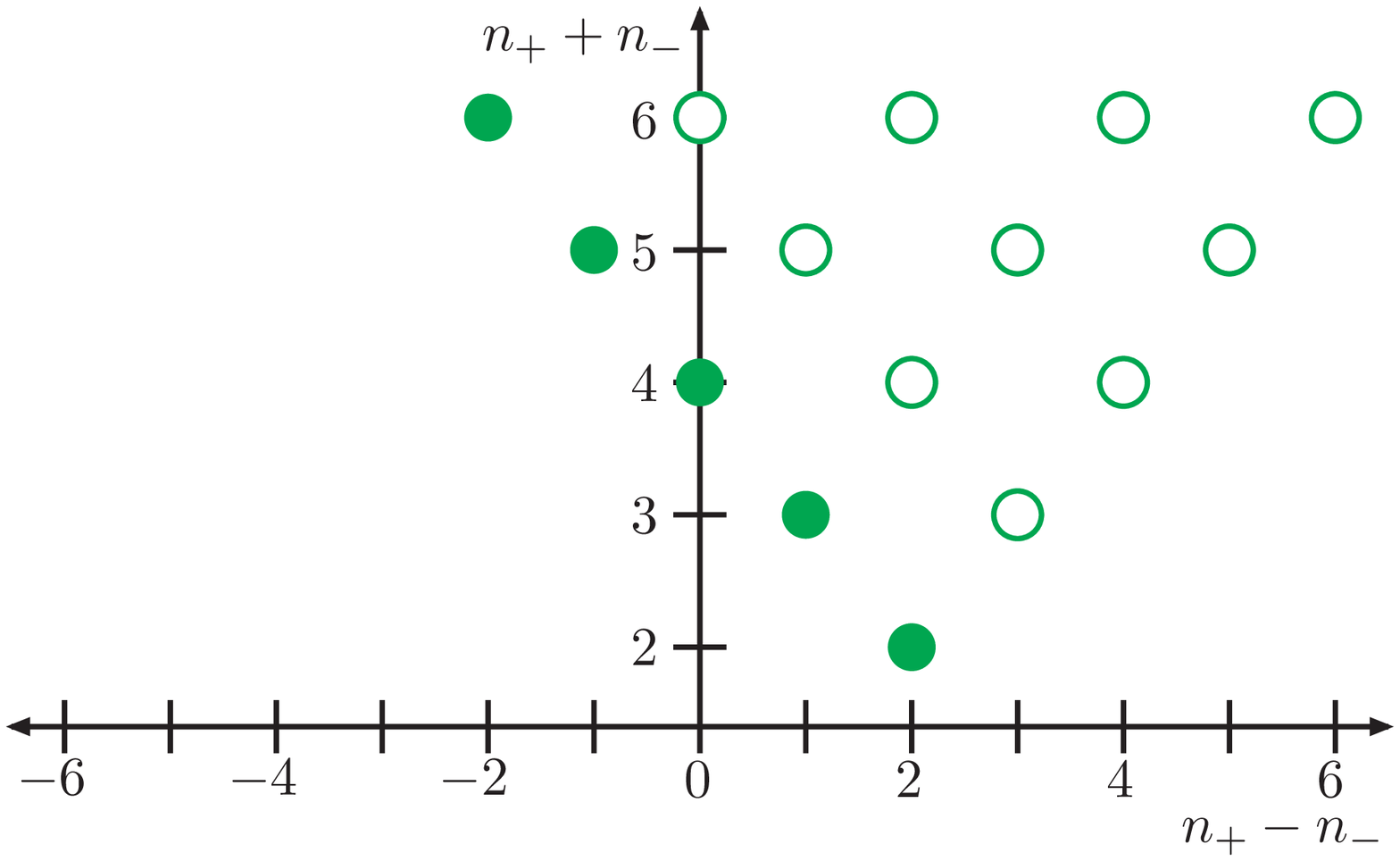}}
\caption{The anti-MHV structure of
$\phi^\dagger$ plus multi-gluon amplitudes.
The axes are as in \fig{PhiTreeMapFigure}.
Solid green dots represent fundamental `$\phi^\dagger$-anti-MHV' vertices,
which coincide with the $\phi^\dagger g^+g^+ g^- \ldots g^-$ amplitudes.
Open green circles represent $\phi^\dagger$ amplitudes which are composites,
built from the  $\phi^\dagger$-anti-MHV vertices plus pure-gauge-theory anti-MHV
vertices. These amplitudes can also be obtained from the $\phi$ amplitudes 
by parity, which exchanges $\spa{i}.{j} \lr \spb{j}.{i}$ and reflects 
points across the vertical axis, $n_+-n_- \to -(n_+-n_-)$.
}
\label{PhibarTreeMapFigure}
}
%%%%%%%%%%%%%%%%%%%%

The corresponding amplitudes for $\phi^\dagger$ are shown in
\fig{PhibarTreeMapFigure}. They can be obtained by applying parity to the
$\phi$ amplitudes. For practical purposes this means that we compute with
$\phi$, and reverse the helicities of every gluon. Then we let
$\spa{i}.{j} \lr \spb{j}.{i}$ to get the desired $\phi^\dagger$
amplitude. The set of building-block amplitudes are therefore anti-MHV.
Furthermore, the amplitudes with additional positive-helicity gluons are
obtained by combining with anti-MHV gauge theory vertices. The
anti-MHV-drift is always to the right and upwards.  Collectively, these
amplitudes form the anti-holomorphic (or anti-MHV) tower of accessible
amplitudes.

%%%%%%%%%%%%%%%%%%%%
%FIGURE
%
\FIGURE[t]{
{\epsfxsize 4 truein \epsfbox{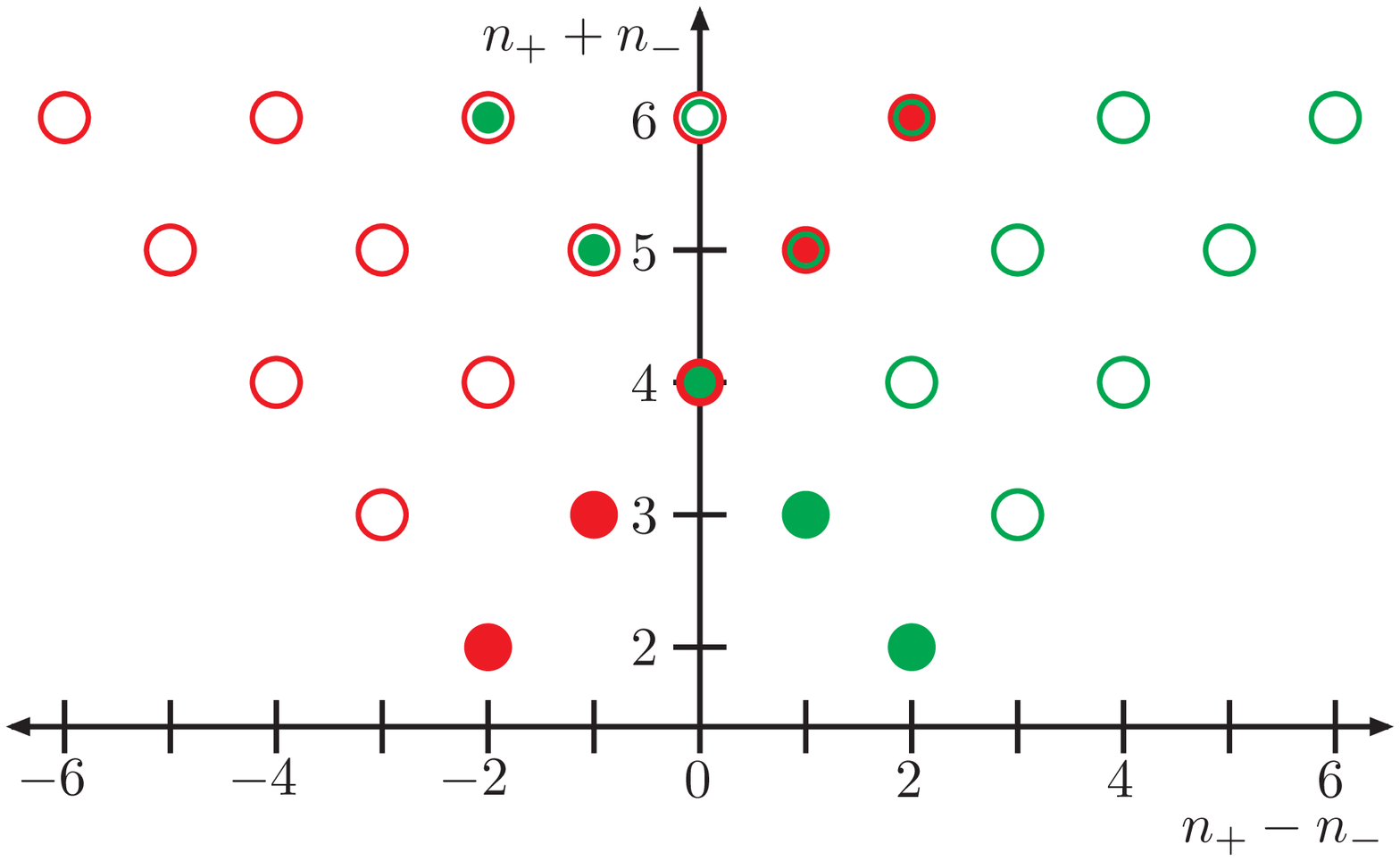}}
\caption{The structure of Higgs plus multi-gluon amplitudes obtained by
combining the MHV tower for $\phi+n$~gluons and the anti-MHV tower
of $\phi^\dagger +n$~gluon amplitudes.
The axes are as in \fig{PhiTreeMapFigure}.
Note that the point at $n_+ + n_- = 2$, $n_+ - n_- = 0$, $H \to g^+ g^-$,
vanishes by angular momentum conservation.
The MHV tower from \fig{PhiTreeMapFigure} is shown in red.
The anti-MHV tower from \fig{PhibarTreeMapFigure} is shown in green.
Amplitudes for the scalar Higgs are obtained by adding the $\phi$ and
$\phi^\dagger$ amplitudes.}
\label{TreeMapFigure}
}
%%%%%%%%%%%%%%%%%%%%%%

To obtain amplitudes for the real Higgs boson with gluons,
we merely add the $\phi$ and $\phi^\dagger$ amplitudes.  
The allowed helicity states are shown in \fig{TreeMapFigure} 
and are composed of both holomorphic and anti-holomorphic structures.

%%%%%%%%%%%%%%%%%%%%%%%%%%%%%%%%%%%

\section{MHV rules and applications}
\label{MHVRulesSection}

The new MHV rules for computing Higgs plus $n$-gluon scattering 
amplitudes can be summarized as follows:
\begin{enumerate}
\item For the $\phi$ couplings, everything is exactly like CSW~\cite{CSW1}
(except for the momentum carried by $\phi$).
\item For $\phi^\dagger$, we just apply parity.  That is, we compute with $\phi$,
and reverse the helicities of every gluon.
Then we let $\spa{i}.{j} \lr \spb{j}.{i}$ to get the desired $\phi^\dagger$
amplitude.
\item For $H$, we add the $\phi$ and $\phi^\dagger$ amplitudes.
\end{enumerate}

These rules generate a set of amplitudes with all the correct collinear
and multi-particle factorization properties, as follows from
the same type of argument as in the pure gauge theory case~\cite{CSW1}.
In addition, the rules can easily be used to reproduce all of the 
available analytic formulae for Higgs $+$ $n$-gluon scattering ($n\leq5$) 
at tree level. (In some cases the agreement was checked numerically.)
In some instances they generate considerably shorter expressions.
To make things even more efficient, one can use a recursive version 
of the rules, along the lines suggested by ref.~\cite{BBK}.

As a by-product, we also obtain the amplitudes for a pseudoscalar Higgs
boson $A$ plus multiple gluons, where $A$ couples to gluons via
the $A \, G_{\mu\nu}\,{}^*G^{\mu\nu}$ interaction in \eqn{effinta}.
The minimal supersymmetric Standard Model contains such a field.
If $A$ is light enough, and the top quark dominates the loop,
this effective interaction is a good approximation.
Then to construct $A_n(A,1,2,\ldots,n)$, in step 3 of the rules 
we merely need to take the difference instead of the sum of the 
corresponding $\phi$ and $\phi^\dagger$ amplitudes.

For the $\phi$ plus $n$-gluon amplitudes, we can consider a
twistor space $(\lambda_1,\lambda_2,\mu^{\dot1},\mu^{\dot2})$, 
for each of the $n$ gluons~\cite{Witten1}, by replacing
the anti-holomorphic spinor coordinates $\tilde\lambda_{i,\dot\alpha}$
by their Fourier transforms,
$\mu_i^{\dot\alpha} = -i\partial/\partial\tilde\lambda_{i,\dot\alpha}$.
In doing this tranformation, we leave the momentum of the massive 
scalar untouched.  Then the argument that the MHV QCD amplitudes are 
localized on a line~\cite{Witten1} extends trivially to the $\phi$
amplitudes~(\ref{assrtn}).   
The recoiling momentum of the $\phi$ particle enters the
overall momentum-conserving delta function, but this does not affect the
localization of the Fourier transform,
\bea 
A(\lambda_i,\mu_i) 
&=& \int \prod_{i=1}^n d\tilde\lambda_i \exp(i\mu_i\tilde\lambda_i)
   A(\lambda_i) \delta\Bigl( k_\phi + \sum_{i=1}^n k_i \Bigr) 
\nonumber \\
&=& \int d^4 x A(\lambda_i)
 \int \prod_{i=1}^n d\tilde\lambda_i
\exp(i\mu_i\tilde\lambda_i)
 \exp\Bigl[ i x^{\dot\alpha\alpha} \Bigl( 
  (k_\phi)_{\alpha\dot\alpha} 
  + \sum_{i=1}^n \lambda_{i,\alpha} \tilde\lambda_{i,\dot\alpha}
  \Bigr) \Bigr] \,,
\nonumber \\
&=& \int d^4 x \exp( i x \cdot k_\phi) A(\lambda_i)
    \prod_{i=1}^n \delta( \mu_i + x \lambda_i )\,.
\label{TwistorDelta}
\eea
The $\phi$ amplitudes with $n_-$ negative-helicity gluons are 
similarly localized on networks of $(n_- - 1)$ intersecting lines
in twistor space.

%%%%%%%%%%%%%%%%

\subsection{NMHV amplitudes $A_n(\phi,\ldots,m_1^-,\ldots,m_2^-,\ldots,m_3^-,\ldots)$}

We start by deriving the Next-to-MHV (NMHV) amplitude 
$A_n(\phi,m_1^-,m_2^-,m_3^-)$.
{}From now on we will suppress the dots for positive-helicity gluons in the
MHV tower of amplitudes.
The two topologically distinct diagrams are shown in 
%\fig{NMHVFigure}.  
figure 4. 
Each of these diagrams is drawn for a fixed arrangement of negative-helicity
gluons, such that
$\phi$ is followed by $m_1$.
To obtain the full NMHV amplitude we need to sum over the three cyclic permutations
of $m_1$, $m_2$ and $m_3$, denoted by $C(m_1,m_2,m_3)$.
The full NMHV amplitude is given by,
\begin{equation}
\label{eq:NMHV}
A_n(\phi,m_1^-,m_2^-,m_3^-) = \, \frac{1}{\prod_{l=1}^n \spa{l,}.{l+1}}\,
\sum_{i=1}^2 
\sum_{C(m_1,m_2,m_3)}
A_n^{(i)} (m_1,m_2,m_3) \ ,
\end{equation}
where the common standard denominator is factored out for convenience.
We label the gluon momenta as $k_i$ (where $i$ is defined modulo 
$n$) and introduce the composite (off-shell) momentum,
\begin{equation}
q_{i+1,j} = k_{i+1} + k_{i+2} + \cdots + k_j \,.
\end{equation}
Note that the momentum of $\phi$, $k_\phi$, does not
enter the sum. In particular, $q_{i+1,i} = - k_{\phi}$.
As usual, the  off-shell continuation of the helicity spinor is defined as
\cite{CSW1},
\begin{equation}
(\lambda_{i+1,j})_\alpha = (q_{i+1,j})_{\alpha\dot\alpha} \,\xi^{\dot\alpha},
\end{equation}
where $\xi^{\dot\alpha}$ is a reference spinor that can be chosen
arbitrarily.

\begin{figure}[t]
\label{NMHVFigure}
\psfrag{i+}{\Huge$i\,+$}
\psfrag{i+1+}{\Huge$(i+1)\,+$}
\psfrag{j+}{\Huge$j\,+$}
\psfrag{j+1+}{\Huge$(j+1)\,+$}
\psfrag{k+}{\Huge$k\,+$}
\psfrag{n+}{\Huge$n\,+$}
\psfrag{n+}{\Huge$n\,+$}
\psfrag{1-}{\Huge${m_1}\,-$}
\psfrag{2-}{\Huge$2\,-$}
\psfrag{phi}{\Huge$\phi$}
\psfrag{m1-}{\Huge$m_1\,-$}
\psfrag{m2-}{\Huge$m_2\,-$}
\psfrag{m3-}{\Huge$m_3\,-$}
\psfrag{4}{\Huge$4\,+$}
\psfrag{+}{\Huge$+$}
\psfrag{-}{\Huge$-$}
\begin{center}
{\scalebox{0.40}{
\includegraphics{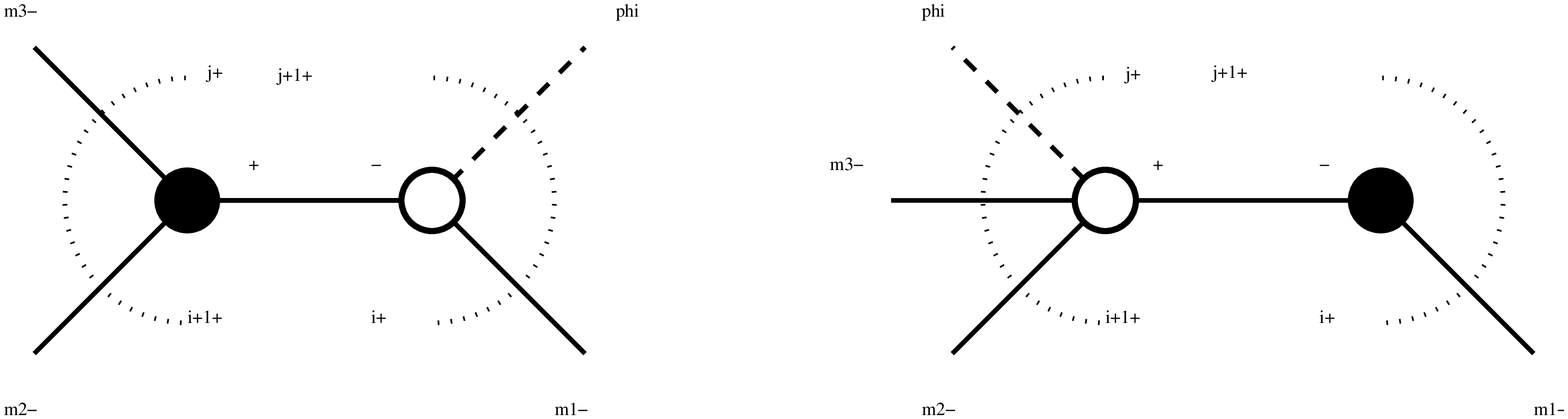}}
}
\end{center}
\caption{ Tree diagrams with MHV vertices which contribute to the
NMHV amplitude
$A_n(\phi,\ldots,m_1^-,\ldots,m_2^-,\ldots,m_3^-,\ldots)$.
The scalar $\phi$ is represented by a dashed line and
negative-helicity gluons, $g^-$,  by solid lines. Positive-helicity gluons
$g^+$ emitted from each vertex are indicated by dotted semicircles, 
with labels showing the bounding $g$ lines in each MHV vertex. }
\end{figure}

Following the organisational structure of ref.~\cite{GGK},
the contributions of the individual diagrams in figure 4 are,
\begin{eqnarray}
A_n^{(1)}(m_1,m_2,m_3) &=& 
\sum_{i=m_1}^{m_2-1} \sum_{j=m_3}^{m_1-1}
\frac{ \spa{m_2}.{m_3}^4 \langle m_1^- | \slash\!\!\! q_{i+1,j} |
\xi^-\rangle^4}{D(i,j,q_{i+1,j})} \,,\nonumber \\
A_n^{(2)}(m_1,m_2,m_3) &=& 
\sum_{i=m_1}^{m_2-1} \sum_{j=m_3}^{m_1-1}
\frac{ \spa{m_2}.{m_3}^4
\langle m_1^- | \slash\!\!\! q_{j+1,i} | \xi^-\rangle^4}{D(i,j,q_{j+1,i})}
\,,
\end{eqnarray}
where
\begin{equation}
\label{eq:Ddef}
D(i,j,q) =
\langle i^- | \slash\!\!\! q |\xi^-\rangle
\langle (j+1)^- | \slash\!\!\! q |\xi^-\rangle
\langle (i+1)^- | \slash\!\!\! q |\xi^-\rangle
\langle j^- | \slash\!\!\! q |\xi^-\rangle
\frac{q^2}{\spa{i,}.{i+1}\spa{j,}.{j+1}} \,.
\end{equation}
In the summation over $j$, it should be understood that the maximum value
is taken modulo $n$.  In other words, when $m_1=1$, the upper limit is 
$0 \equiv n$, but when $m_1=2$, the upper limit is 1.  Note that diagrams of
the second type vanish when there are no positive-helicity gluons emitted
from the right hand vertex.  In this case, $j+1 = m_1 = i$ and $q_{j+1,i}
= q_{m_1}$.  These diagrams are automatically killed by the $\langle m_1^-
| q_{j+1,i}$ factor in the numerator.

In distinction with ref.~\cite{CSW1}, we leave the reference 
spinor $\xi$ arbitrary and specifically do not set it to be
equal to one of the momenta in the problem.
This has two advantages.   First, we do not introduce
unphysical singularities for gluonic amplitudes (for generic
points in phase space);
and second, it allows a powerful numerical check of gauge
invariance (which all of our amplitudes satisfy).

The amplitude~\ref{eq:NMHV} describes all amplitudes
coupling $\phi$ to 3 negative-helicity gluons and
any number of positive-helicity gluons.
In particular, it describes the $\phi \to {-}{-}{-}$ and 
$\phi\to {+}{-}{-}{-}$ amplitudes.  These amplitudes only receive contributions
from the MHV tower of amplitudes and are therefore also the amplitudes for
$H \to {-}{-}{-}$ and $H \to {+}{-}{-}{-}$.

%%%%%%%%%%%%%%%%

\subsubsection{$H \to {-}{-}{-}$}

In this case, we can take $m_1=1$, $m_2 = 2$ and $m_3=3$.  
The second class of diagrams in figure 4 collapses since there are
not enough gluons to prevent the right hand vertex vanishing. Hence
$A_3^{(2)}=0$.  From the first class of diagrams, $A_3^{(1)}$,
three individual diagrams survive.  (Our counting includes the
cyclic permutations of $m_1$, $m_2$ and $m_3$, as required 
in \eqn{eq:NMHV}.)

%%%%%%%%%%%%%%%%

\subsubsection{$H \to {+}{-}{-}{-}$}

In this case, we can take $m_1=2$, $m_2 = 3$ and $m_3=4$.
Seven individual diagrams survive, as can be seen from figure 4, 
including appropriate cyclic permutations.
Five of them are of the first type, $A_4^{(1)}$,
where $\phi$ couples directly to one on-shell negative-helicity gluon; 
two are of the second type, $A_4^{(2)}$, where
$\phi$ couples directly to two negative-helicity gluons.

In both cases, we have checked, 
with a help of a symbolic manipulator, that our results are $\xi$-independent
(gauge invariant) and agree numerically with the known analytic formulae,
\begin{eqnarray}
A_3(H,1^-,2^-,3^-)&=& - \frac{m_H^4}{\spb{1}.{2}\spb{2}.{3}\spb{3}.{1}}\,,
\\
A_4(H,1^+,2^-,3^-,4^-)&=& 
\frac{\langle 3^-|\slash\!\!\! k_H | 1^- \rangle^2 \spa{2}.{4}^2}
{s_{124} s_{12} s_{14}}
+\frac{\langle 4^-|\slash\!\!\! k_H | 1^- \rangle^2 \spa{2}.{3}^2}
{s_{123} s_{12} s_{23}} 
+\frac{\langle 2^-|\slash\!\!\! k_H | 1^- \rangle^2 \spa{3}.{4}^2}
{s_{134} s_{14} s_{34}}
\nonumber\\
&& \hskip-0.6cm 
-
\frac{\spa{2}.{4}}{\spa{1}.{2}\spb{2}.{3}\spb{3}.{4}\spa{4}.{1}}
\left(s_{23}\frac{\langle 2^-|\slash\!\!\! k_H | 1^- \rangle}{\spb{4}.{1}}
+s_{34}\frac{\langle 4^-|\slash\!\!\! k_H | 1^- \rangle}{\spb{1}.{2}}
- s_{234} {\spa{2}.{4}}
\right) \,. \nonumber \\
\end{eqnarray}
where $k_H = k_\phi$.

%%%%%%%%%%%%%%%%%

\subsection{NNMHV amplitudes $A_n(\phi,\ldots,m_1^-,\ldots,m_2^-,\ldots,m_3^-,
\ldots,m_4^-,\ldots)$}

The Next-to-Next-to-MHV
(NNMHV) amplitudes follow from diagrams with three MHV vertices. 
The skeleton diagram in figure 5 shows our labelling conventions
for gluons.

\begin{figure}
\label{fig:skeleton}
\psfrag{i+}{\Large$i\,+$}
\psfrag{i+1+}{\Large$(i+1)\,+$}
\psfrag{j+}{\Large$j\,+$}
\psfrag{j+1+}{\Large$(j+1)\,+$}
\psfrag{k+}{\Large$k\,+$}
\psfrag{k+1+}{\Large$(k+1)\,+$}
\psfrag{l+}{\Large$l\,+$}
\psfrag{l+1+}{\Large$(l+1)\,+$}
\begin{center}
{\scalebox{0.60}{
\includegraphics{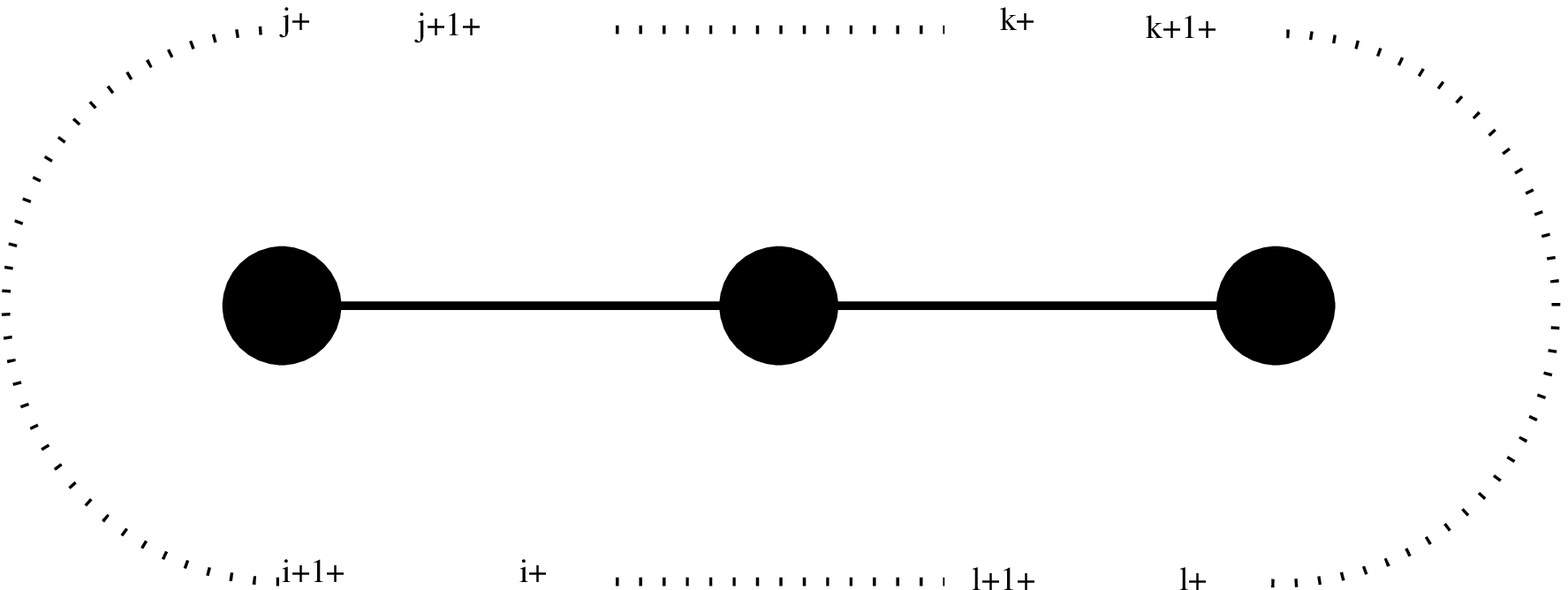}}
}
\end{center}
\caption{Skeleton diagram showing the labelling of $n$  
gluons for amplitudes $A_n$ with four negative-helicity gluons.
Positive-helicity gluons
$g^+$ emitted from each vertex are indicated by dotted lines with labels
showing the bounding $g^+$ lines in each MHV vertex.}
\end{figure}

There are thirteen topologically distinct diagrams in this case, shown in 
%\fig{NNMHV}.  
figure 6. 
As before, each of these diagrams is drawn for the fixed arrangement 
of negative-helicity gluons, such that $\phi$ is followed by $m_1$.
To obtain the full NNMHV amplitude we need to sum over all cyclic permutations,
$C(m_1,m_2,m_3,m_4)$.
The resulting total amplitude is given by,
\begin{equation}
\label{eq:NNMHV}
A_n(\phi,m_1^-,m_2^-,m_3^-,m_4^-) =\, \frac{1}{\prod_{l=1}^n \spa{l,}.{l+1}}\,
\sum_{i=1}^{13}
\sum_{C(m_1,m_2,m_3,m_4)}
A_n^{(i)} (m_1,m_2,m_3,m_4) \ .
\end{equation}

The contributions of the first five diagrams in figure 6 are,
\begin{eqnarray}
A_n^{(1)}(m_1,m_2,m_3,m_4) &=& 
\sum_{k=m_4}^{m_1-1} \sum_{j=m_4}^{k} \sum_{i=m_2}^{m_3-1}  
\sum_{l=m_1}^{m_2-1} 
\frac{ \spa{m_3}.{m_4}^4 
\langle m_2^- | \slash\!\!\! q_{i+1,j} |\xi^-\rangle^4
\langle m_1^- | \slash\!\!\! q_{l+1,k} |\xi^-\rangle^4}
{DD(i,j,q_{i+1,j},k,l,q_{l+1,k})} \,, \nonumber \\
A_n^{(2)}(m_1,m_2,m_3,m_4) &=& 
\sum_{k=m_4}^{m_1-1} \sum_{j=m_3}^{m_4-1} \sum_{i=m_1}^{m_2-1}  
\sum_{l=m_1}^{i} 
\frac{ \spa{m_2}.{m_3}^4 
\langle m_4^- | \slash\!\!\! q_{i+1,j} |\xi^-\rangle^4
\langle m_1^- | \slash\!\!\! q_{l+1,k} |\xi^-\rangle^4}
{DD(i,j,q_{i+1,j},k,l,q_{l+1,k})} \,, \nonumber \\
A_n^{(3)}(m_1,m_2,m_3,m_4) &=&
\sum_{k=m_4}^{m_1-1} \sum_{j=m_4}^{k} \sum_{i=m_3}^{m_4-1}  
\sum_{l=m_1}^{m_2-1} 
\frac{ \spa{m_2}.{m_3}^4 
\langle m_4^- | \slash\!\!\! q_{i+1,j} |\xi^-\rangle^4
\langle m_1^- | \slash\!\!\! q_{l+1,k} |\xi^-\rangle^4}
{DD(i,j,q_{i+1,j},k,l,q_{l+1,k})} \,, \nonumber \\
A_n^{(4)}(m_1,m_2,m_3,m_4) &=& 
\sum_{k=m_4}^{m_1-1} \sum_{j=m_2}^{m_3-1} \sum_{i=m_1}^{m_2-1}  
\sum_{l=m_1}^{i}
\frac{ \spa{m_3}.{m_4}^4 
\langle m_2^- | \slash\!\!\! q_{i+1,j} |\xi^-\rangle^4
\langle m_1^- | \slash\!\!\! q_{l+1,k} |\xi^-\rangle^4}
{DD(i,j,q_{i+1,j},k,l,q_{l+1,k})} \,, \nonumber \\
A_n^{(5)}(m_1,m_2,m_3,m_4) &=& 
\sum_{k=m_4}^{m_1-1} \sum_{j=m_3}^{m_4-1} \sum_{i=m_2}^{m_3-1}  
\sum_{l=m_1}^{m_2-1} 
\frac{ \spa{m_2}.{m_4}^4 
\langle m_3^- | \slash\!\!\! q_{i+1,j} |\xi^-\rangle^4
\langle m_1^- | \slash\!\!\! q_{l+1,k} |\xi^-\rangle^4}
{DD(i,j,q_{i+1,j},k,l,q_{l+1,k})} \,. \nonumber \\
\label{first5}
\end{eqnarray}

\begin{figure}[th!]
\label{fig:NNMHV}
\psfrag{i+}{\Huge$i\,+$}
\psfrag{i+1+}{\Huge$(i+1)\,+$}
\psfrag{j+}{\Huge$j\,+$}
\psfrag{j+1+}{\Huge$(j+1)\,+$}
\psfrag{k+}{\Huge$k\,+$}
\psfrag{n+}{\Huge$n\,+$}
\psfrag{n+}{\Huge$n\,+$}
\psfrag{1-}{\Huge${m_1}\,-$}
\psfrag{2-}{\Huge$2\,-$}
\psfrag{m1-}{\Huge$m_1^-$}
\psfrag{m2-}{\Huge$m_2^-$}
\psfrag{m3-}{\Huge$m_3^-$}
\psfrag{m4-}{\Huge$m_4^-$}
\psfrag{phi}{\Huge$\phi$}
\psfrag{4}{\Huge$4\,+$}
\psfrag{+}{\Huge$+$}
\psfrag{-}{\Huge$-$}
\begin{center}
{\scalebox{0.25}{
\includegraphics{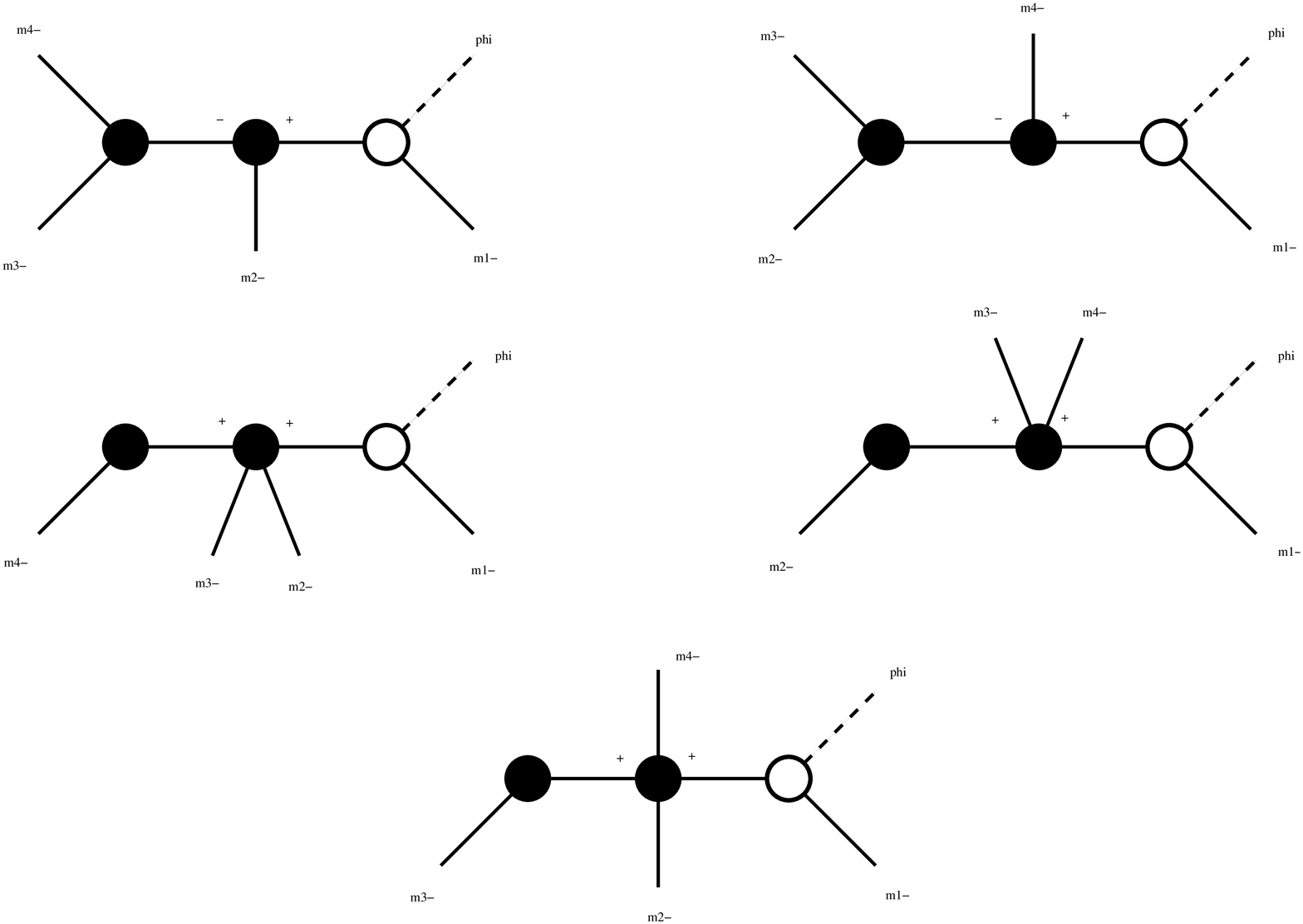}}
}
{\scalebox{0.25}{
\includegraphics{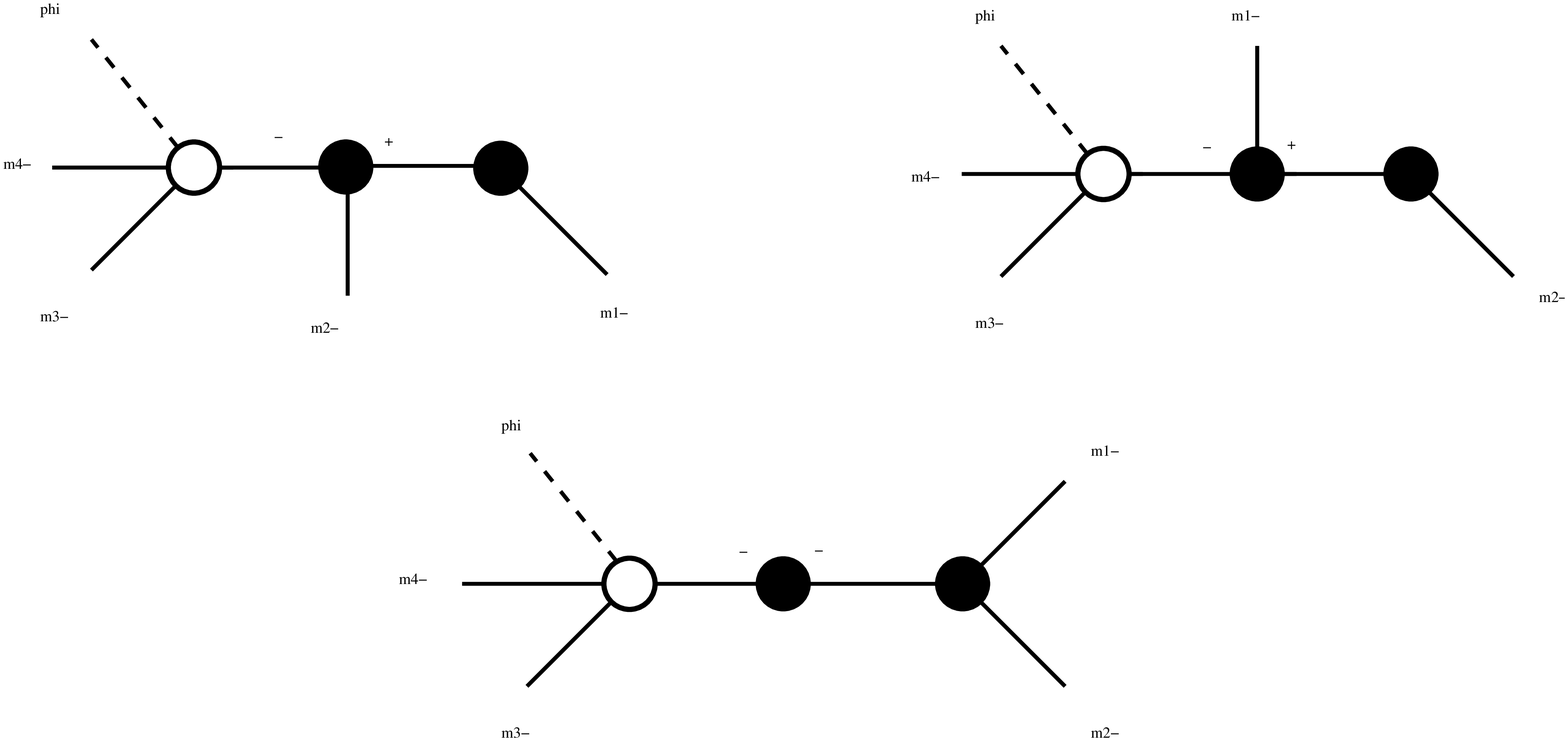}}
}
{\scalebox{0.25}{
\includegraphics{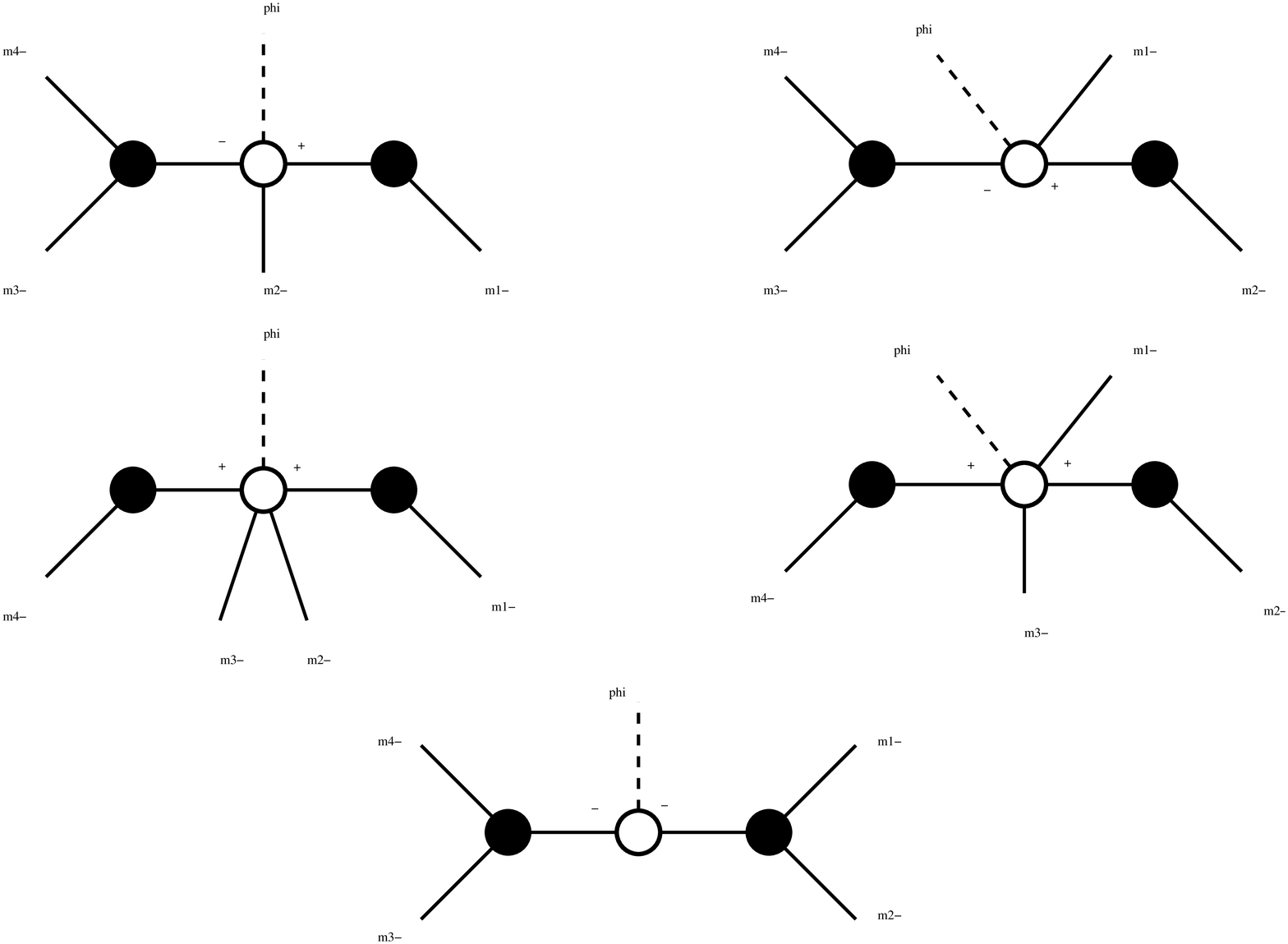}}
}
\end{center}
\caption{NNMHV tree diagrams contributing to the
amplitude $A_n(\phi,g_{m_1}^-,g_{m_2}^-,g_{m_3}^-,g_{m_4}^-)$.}
\end{figure} 
%\newpage

A comment is in order concerning the boundary values in the sums over
external gluons in \eqn{first5} for cases where it is possible to have no
external gluon legs emitted from a given vertex (and in a given
range). Consider, for example, the first diagram in figure 6.  There are
no negative-helicity gluons, $g^-$, emitted upwards from the middle
vertex, and the number of $g^+$ gluons, $n_+$, can be zero or non-zero.
The summations over $j$ and $k$ in $A_n^{(1)}$ should read:
\be 
\sum_{k=m_4}^{m_1-1} \, \sum_{j=m_4}^k
 = 
\sum_{k=m_4+1}^{m_1-1} \, \sum_{j=m_4}^{k-1} \, + \,  
 \sum_{k=m_4}^{m_1-1} \, \bigg|_{j=k} 
  \,,     
\ee
where the first term (the double sum) corresponds to $n_+>0$,
and the second term takes into account the case of
$n_+=0$, or no $g^+$ gluons emitted upwards from the middle vertex.

The effective propagator $DD$ is defined by,
\begin{equation}
DD(i,j,q_1,k,l,q_2) = \chi(j,k,q_1,q_2)\chi(l,i,q_2,q_1)
\, D(i,j,q_1)\, D(k,l,q_2) \,,
\end{equation}
where $D$ is given in \eqn{eq:Ddef}, and where
\begin{eqnarray}
\chi(j,k,q_1,q_2) &=& 1 
\qquad\qquad\qquad\qquad\qquad~{\rm ~~~if~} j \neq k, \nonumber \\
&=&\frac{
\spa{j,}.{j+1}
\langle \xi^+ | \slash\!\!\! q_1 \slash\!\!\! q_2 |\xi^-\rangle}
{ \langle (j+1)^-| \slash\!\!\! q_1 |\xi^-\rangle
  \langle j^- |  \slash\!\!\! q_2 |\xi^-\rangle}
  {\rm ~~~if~} j = k.
\end{eqnarray}

The contributions of diagrams 6, 7 and 8 in figure 6 read,
\begin{eqnarray}
A_n^{(6)}(m_1,m_2,m_3,m_4) &=& 
\sum_{k=m_4}^{m_1-1} \sum_{j=m_4}^{k} \sum_{i=m_2}^{m_3-1} 
\sum_{l=m_1}^{m_2-1} 
\frac{ \spa{m_3}.{m_4}^4 
\langle m_2^- | \slash\!\!\! q_{j+1,i} |\xi^-\rangle^4
\langle m_1^- | \slash\!\!\! q_{k+1,l} |\xi^-\rangle^4}
{DD(k,l,q_{k+1,l},i,j,q_{j+1,i})} \,, \nonumber \\
A_n^{(7)}(m_1,m_2,m_3,m_4) &=& 
\sum_{k=m_1}^{m_2-1} \sum_{j=m_4}^{m_1-1} \sum_{i=m_2}^{m_3-1} 
\sum_{l=m_2}^{i} 
\frac{ \spa{m_3}.{m_4}^4 
\langle m_1^- | \slash\!\!\! q_{j+1,i} |\xi^-\rangle^4
\langle m_2^- | \slash\!\!\! q_{k+1,l} |\xi^-\rangle^4}
{DD(k,l,q_{k+1,l},i,j,q_{j+1,i})}  \,, \nonumber \\
A_n^{(8)}(m_1,m_2,m_3,m_4) &=& 
\sum_{k=m_4}^{m_1-1} \sum_{j=m_4}^{k} \sum_{i=m_2}^{m_3-1} 
\sum_{l=m_2}^{i} 
\frac{ \spa{m_3}.{m_4}^4 
\langle \xi^+ | \slash\!\!\! q_{j+1,i} 
\, \slash\!\!\! q_{k+1,l} |\xi^-\rangle^4
\spa{m_1}.{m_2}^4 }
{DD(k,l,q_{k+1,l},i,j,q_{j+1,i})} \,. \nonumber \\
\end{eqnarray}

Finally, for the last five diagrams in figure 6 we have,
\begin{eqnarray}
A_n^{(9)}(m_1,m_2,m_3,m_4) &=&
\sum_{k=m_4}^{m_1-1} \sum_{j=m_4}^{k} \sum_{i=m_2}^{m_3-1} 
\sum_{l=m_1}^{m_2-1}
\frac{ \spa{m_3}.{m_4}^4 
\langle m_2^- | \slash\!\!\! q_{i+1,j} |\xi^-\rangle^4
\langle m_1^- | \slash\!\!\! q_{k+1,l} |\xi^-\rangle^4}
{DD(i,j,q_{i+1,j},k,l,q_{k+1,l})} \,, \nonumber \\
A_n^{(10)}(m_1,m_2,m_3,m_4) &=& 
\sum_{k=m_1}^{m_2-1}\sum_{j=m_4}^{m_1-1} \sum_{i=m_2}^{m_3-1} 
\sum_{l=m_2}^{i} 
\frac{ \spa{m_3}.{m_4}^4 
\langle m_1^- | \slash\!\!\! q_{i+1,j} |\xi^-\rangle^4
\langle m_2^- | \slash\!\!\! q_{k+1,l} |\xi^-\rangle^4}
{DD(i,j,q_{i+1,j},k,l,q_{k+1,l})} \,, \nonumber \\
A_n^{(11)}(m_1,m_2,m_3,m_4) &=& 
\sum_{k=m_4}^{m_1-1} \sum_{j=m_4}^{k} \sum_{i=m_3}^{m_4-1} 
\sum_{l=m_1}^{m_2-1}
\frac{ \spa{m_2}.{m_3}^4 
\langle m_4^- | \slash\!\!\! q_{i+1,j} |\xi^-\rangle^4
\langle m_1^- | \slash\!\!\! q_{k+1,l} |\xi^-\rangle^4}
{DD(i,j,q_{i+1,j},k,l,q_{k+1,l})} \,, \nonumber \\
A_n^{(12)}(m_1,m_2,m_3,m_4) &=& \frac{1}{2}
\sum_{k=m_1}^{m_2-1} \sum_{j=m_4}^{m_1-1} \sum_{i=m_3}^{m_4-1} 
\sum_{l=m_2}^{m_3-1} 
\frac{ \spa{m_1}.{m_3}^4 
\langle m_4^- | \slash\!\!\! q_{i+1,j} |\xi^-\rangle^4
\langle m_2^- | \slash\!\!\! q_{k+1,l} |\xi^-\rangle^4}
{DD(i,j,q_{i+1,j},k,l,q_{k+1,l})} , \nonumber \\
A_n^{(13)}(m_1,m_2,m_3,m_4) &=& \frac{1}{2}
\sum_{k=m_4}^{m_1-1} \sum_{j=m_4}^{k} \sum_{i=m_2}^{m_3-1} 
\sum_{l=m_2}^{i}
\frac{ \spa{m_3}.{m_4}^4 
\langle \xi^+ | \slash\!\!\! q_{i+1,j} 
\, \slash\!\!\! q_{k+1,l} |\xi^-\rangle^4
\spa{m_1}.{m_2}^4 }
{DD(i,j,q_{i+1,j},k,l,q_{k+1,l})} \,. \nonumber \\
\end{eqnarray}
Note that the expressions for the last two diagrams, $A_n^{(12)}$ and 
$A_n^{(13)}$, contain a factor of $\hf$. This factor is necessary to 
take into account the fact that, out of the four cyclic permutations 
$C(m_1,m_2,m_3,m_4)$ in \eqn{eq:NNMHV}, only two give inequivalent 
diagrams; the remaining two double up the result.

%%%%%%%%%%%%%%%

\subsubsection{$H \to {-}{-}{-}{-}$}

Our general NNMHV expressions \eqn{eq:NNMHV} can be applied to the simple
case with no positive-helicity gluons, $n_+=0$.  Only diagrams 1, 2 and 13
survive --- all others give zero contribution because there are not enough
gluons to prevent one of the vertices from vanishing.  We checked
numerically that our result is gauge invariant and agrees with the known
expression,
\be
A_4(H,1^-,2^-,3^-,4^-) = A_4(\phi,1^-,2^-,3^-,4^-) =
 \frac{m_H^4}{\spb{1}.{2}\spb{2}.{3}\spb{3}.{4}\spb{4}.{1}} \,.
\ee

%%%%%%%%%%%

\subsubsection{$H \to {+}{-}{-}{-}{-}$}

The amplitude for $n_+=1$,
\be
A_5(H,1^+,2^-,3^-,4^-,5^-) = A_5(\phi,1^+,2^-,3^-,4^-,5^-) \,,
\ee
can be obtained by setting $m_1 = 2$, $m_2=3$, $m_3=4$ and $m_4=5$ in
\eqn{eq:NNMHV}. The result is again gauge invariant, 
and we have checked that it agrees numerically with eq.~(B.2) 
in ref.~\cite{DFM}.

%%%%%%%%%%%%

\subsection{$H \to {-}{-}{-}\cdots {-}$}

In Appendix~\ref{RecursiveAllPlusSection} we present a recursive 
construction of all $n$-gluon
amplitudes with gluons of the same helicity:
$A_n(\phi^\dagger,1^+,2^+,\ldots,n^+)$, with parity conjugates
$A_n(\phi,1^-,2^-,\ldots,n^-)$.  This derivation makes use of the
Berends-Giele off-shell currents \cite{BerendsGiele}. 

In this Section we will derive the same amplitudes using the MHV rules.
The two derivations will turn out to be almost identical; nevertheless, we
believe that it is instructive to demonstrate the all-orders-in-$n$
agreement between the results derived in the standard approach
(Appendix~\ref{RecursiveAllPlusSection}) and in the MHV rules approach 
(this Section).

\begin{figure}
\label{fig:allplus}
\psfrag{-}{\Huge$-$}
\psfrag{i}{\Huge$i\,-$}
\psfrag{i+1}{\Huge$(i+1)\,-$}
\psfrag{j}{\Huge$j\,-$}
\psfrag{j+1}{\Huge$(j+1)\,-$}
\psfrag{phi}{\Huge$\phi$}
\begin{center}
{\scalebox{0.35}{
\includegraphics{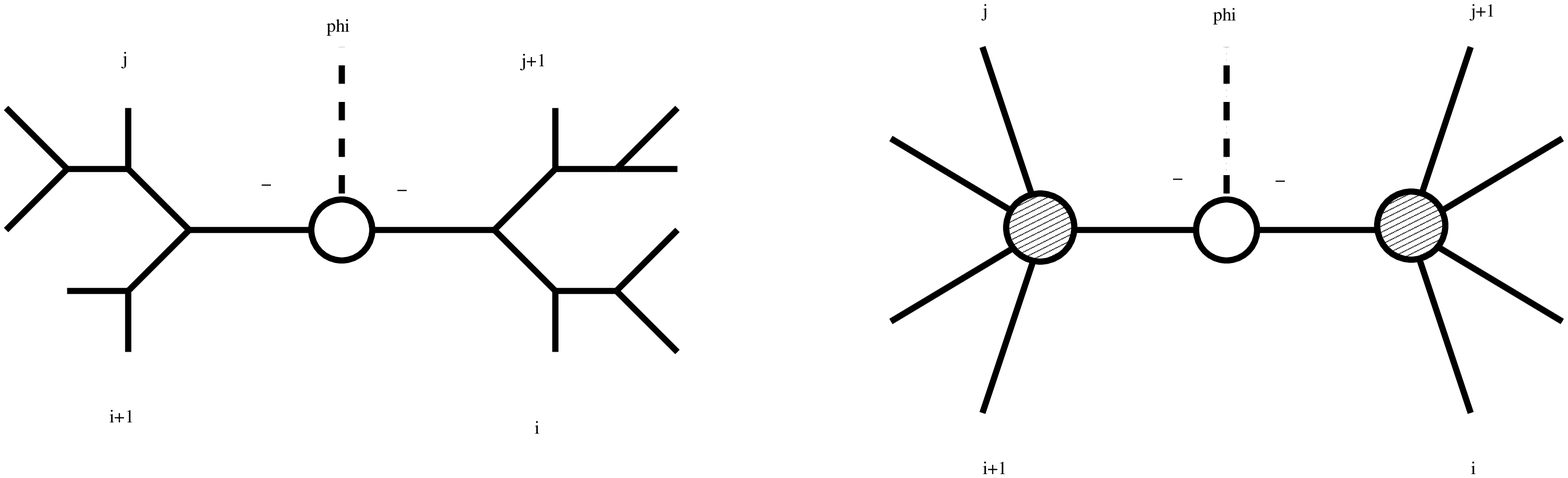}}
}
\end{center}
\caption{Two representations for $A_n(\phi,1^-,2^-,3^-,\ldots,n^-)$
in the MHV-rules approach.  In (a), we illustrate how 
attaching only negative-helicity gluons requires just three-point 
MHV vertices.  In (b), the shaded circle represents the coupling 
of an off-shell gluon to many on-shell negative helicity gluons, 
which is obtained by summing MHV graphs of the type shown in (a).}
\end{figure}

In the MHV-rules approach, the `minus-only' amplitudes,
\be
A_n(H,1^-,2^-,3^-,\ldots,n^-) = A_n(\phi,1^-,2^-,3^-,\ldots,n^-) \,,
\ee
are constructed by attaching $n-2$ of the tree-level three-point MHV vertices,
$A_3({+}{-}{-})$, to the $\phi$-MHV two-gluon vertex,
$A_2(\phi,{-}{-})$, in all possible ways, as depicted in figure 7(a).
Each of the two showers of three-point MHV vertices shown in figure 7
can be represented by an off-shell effective vertex,
\be
V_{n}(g_1^{+*},g_2^-,g_3^-,\ldots,g_n^-)=\,
{p_1^2 \over \spb{\xi}.{2} \spb{n}.{\xi}}\,
{(-1)^n \over \spb{2}.{3}\spb{3}.{4} \cdots \spb{n-1,}.{n}} \,,
\label{Voffsh}
\ee
where it is understood that only the $g^+$ leg is off shell
and hence $p_1^2 \neq 0$.
The expression~(\ref{Voffsh}) was derived in ref.~\cite{Zhu} 
using the MHV rules of CSW \cite{CSW1}, and proved by induction. 
It is also similar in spirit to the all-plus 
Berends-Giele~\cite{BerendsGiele} off-shell current~(\ref{JAllPlus})
used in Appendix~\ref{VanishingSection}.
The factor $(-1)^n$ on the right hand side of
\eqn{Voffsh} reflects our conventions for the 
$\spb{i}.{j}$ spinor 
product, \eqn{Atwo}.

Attaching the off-shell $V$-vertices on both sides of the 
amplitude $A_2(\phi,-,-),$ we get the following expression:
\be
A_n(\phi,1^-,2^-,3^-,\ldots,n^-) =
\, {(-1)^n \over \spb{1}.{2}\spb{2}.{3} \cdots \spb{n}.{1}}\, {\cal A} \,,
\ee
where $ {\cal A}$ is the sum 
\be
{\cal A} = \,-
\sum_{1\le i<j\le n}
{\spb{i,}.{i+1} \spb{j,}.{j+1} 
\over \spb{i}.{\xi} \spb{\xi,}.{i+1} \spb{j}.{\xi} \spb{\xi,}.{j+1}}
\, {\spa{\lambda_{i+1,j}}.{\lambda_{j+1,i}}}^2  \,.
\label{SumA}
\ee
The sum~(\ref{SumA}) (or more precisely, its hermitian conjugate)
is computed in Appendix~\ref{RecursiveAllPlusSection}, in
eqs.~(\ref{NonvanishAllplus1})--(\ref{NonvanishAllplus3}).
The result is $ {\cal A}=m_H^4$.  We conclude that
\be
A_n(\phi,1^-,2^-,3^-,\ldots,n^-) =
\, {(-1)^{n} \, m_H^4\over \spb{1}.{2}\spb{2}.{3} \cdots 
\spb{n}.{1}} \,,
\ee
and, similarly, the parity conjugate amplitude is
\be
A_n(\phi^\dagger,1^+,2^+,3^+,\ldots,n^+) =
\, {m_H^4\over \spa{1}.{2}\spa{2}.{3} \cdots 
\spa{n}.{1}} \,,
\ee
in agreement with the results derived in 
Appendix~\ref{RecursiveAllPlusSection} using standard methods.

%%%%%%%%%%%%

\subsection{Amplitudes in the overlap of two towers}

So far, we have described amplitudes that receive contributions from the
MHV tower of amplitudes. Transcribing these results to amplitudes that lie
in the anti-MHV tower is straightforward. We reverse the helicities of
every gluon and let $\spa{i}.{j} \lr \spb{j}.{i}$ throughout.  For more
complicated objects like 
$\langle m_1^- | \slash \!\!\!\! q | \xi^-\rangle$, 
we make the replacements
\begin{eqnarray}
\langle m_1^- | \,
\slash \!\!\!\! q \,| \xi^-\rangle~&\rightarrow&~\langle m_1^+ |\, 
\slash \!\!\!\! q \,| \xi^+\rangle \,, \nonumber \\
\langle m_1^+ |\, \slash \!\!\!\! q_1 \, \slash \!\!\!\! q_2 \,| \xi^-\rangle%
~&\rightarrow&%
~-\langle m_1^- |\, \slash \!\!\!\! q_1\,\slash \!\!\!\! q_2  \,| \xi^+\rangle
 \,.
\end{eqnarray}

Now we would like to describe amplitudes that lie in the overlap of 
the two towers.

%%%%%%%%%%

\subsubsection{$H \to {+}{+}{-}{-}$}

This is the simplest amplitude which receives contributions from both the
MHV and the anti-MHV towers. The first contribution is simply the
$\phi$-MHV amplitude~(\ref{Hmmpp}), and the second one is its parity
conjugate $\phi^\dagger$-anti-MHV.  In total we have,
\be
 A_4(H,1^+,2^+,3^-,4^-) = { {\spa3.4}^4 \over \spa1.2\spa2.3\spa3.4\spa4.1 }
 +
 { {\spb1.2}^4 \over \spb1.2\spb2.3\spb3.4\spb4.1 } \,,
 \label{Higppmm}
\ee
which is the correct result.

%%%%%%%%%%%%%%%%%

\subsubsection{$H \to {+}{+}{-}{-}{-}$}

This Higgs-plus-five-gluons amplitude receives contributions from both
towers. The contribution from the $\phi$-MHV tower is an NMHV amplitude
$A_5(\phi,1^+,2^+,3^-,4^-,5^-)$ of the type constructed in \eqn{eq:NMHV},
where we set $m_1 = 3$, $m_2=4$ and $m_3=5$.  In total there are 7
contributions of type $A_5^{(1)}$ and 4 of type $A_5^{(2)}$.  The
contribution from the $\phi^\dagger$-anti-MHV tower is the simple anti-MHV
diagram $A_5(\phi^\dagger,1^+,2^+,3^-,4^-,5^-)$, which is the parity
conjugate of \eqn{assrtn},
\be
A_5(\phi^\dagger,1^+,2^+,3^-,4^-,5^-) = 
-{ {\spb1.2}^4 \over \spb1.2\spb2.3\spb3.4\spb4.5\spb5.1 } \,.
\ee
The final result,
\be
A_5(H,1^+,2^+,3^-,4^-,5^-) 
= A_5(\phi,1^+,2^+,3^-,4^-,5^-)
+ A_5(\phi^\dagger,1^+,2^+,3^-,4^-,5^-) \ ,
\label{5ptrest}
\ee
is gauge invariant and agrees numerically with eq.~(B.3) of 
ref.~\cite{DFM}.

\subsection{The soft Higgs limit}

For the case of a massless Higgs boson, we can consider
the kinematic limit where the Higgs momentum goes to zero.
In this limit, because of the form of the $H G_{\mu\nu}G^{\mu\nu}$
interaction, the Higgs boson behaves like a constant,
namely the gauge coupling.  Hence the Higgs-plus-$n$-gluon 
amplitudes should become proportional to the pure-gauge-theory 
amplitudes,
\be
{\cal A}_n(H,\{k_i,\lambda_i,a_i\}) \longrightarrow
({\rm const.}) \times {\del \over \del g} {\cal A}_n(\{k_i,\lambda_i,a_i\}),
\qquad\quad \hbox{as $k_H \to 0$,}
\label{softHiggs1}
\ee
for any helicity configuration.  Taking into account the gauge coupling
factors in the color decomposition~(\ref{TreeColorDecomposition}),
\eqn{softHiggs1} becomes, for the partial amplitudes,
\be
A_n(H,1,2,3,\ldots,n) \longrightarrow 
(n-2) A_n(1,2,3,\ldots,n)
\qquad\quad \hbox{as $k_H \to 0$.}
\label{softHiggs2}
\ee

It is interesting to see how the soft Higgs limit is partitioned
between the $\phi$ and $\phi^\dagger$ amplitudes.   Clearly
the $\phi$-MHV amplitudes~(\ref{assrtn}) become precisely equal
to the corresponding pure-gauge MHV amplitudes~(\ref{MHV}).
{}From figure 4 it is then apparent that the NMHV amplitudes will
approach twice the corresponding pure-gauge MHV amplitudes,
because the field $\phi$ can be attached to either of the two MHV
vertices in the gauge theory case.  From figure 6 there is a factor
of 3 in the NNMHV limit.  More generally,
\be
A_n(\phi,1,2,3,\ldots,n) \longrightarrow 
(n_- - 1) A_n(1,2,3,\ldots,n)
\qquad\quad \hbox{as $k_\phi \to 0$.}
\label{softphi}
\ee
Parity tells us that the $\phi^\dagger$ amplitudes obey,
\be
A_n(\phi^\dagger,1,2,3,\ldots,n) \longrightarrow 
(n_+ - 1) A_n(1,2,3,\ldots,n)
\qquad\quad \hbox{as $k_\phi \to 0$.}
\label{softphidagger}
\ee
Summing \eqns{softphi}{softphidagger} and using $n_+ + n_- = n$,
we recover the soft Higgs limit~(\ref{softHiggs2}).

For $n_- \neq n_+$, the $\phi$ and $\phi^\dagger$ limits are different.
This result is a bit curious, because as $k_\phi \to 0$, the 
$\phi$ and $\phi^\dagger$ interactions become equivalent ---
the $A G_{\mu\nu}\, {}^*G^{\mu\nu}$ coupling they differ by
becomes a total derivative as the field $A$ becomes a constant.
But the interactions are apparently not becoming equivalent fast enough
to prevent the $\phi$ and $\phi^\dagger$ amplitudes from having
different limits.

In the usual MHV-rules approach to amplitudes in pure gauge
theory~\cite{CSW1}, one considers all tree amplitudes to come from
the MHV tower, or sometimes all to come from the anti-MHV-tower.
The $k_H \to 0$ limit of the Higgs amplitude construction suggests
that one can also consider a mixture of the two towers, where
the MHV/anti-MHV content of a given amplitude depends on the
number of positive and negative helicity gluons it contains,
according to \eqns{softphi}{softphidagger}.

%%%%%%%%%%%%%%%%%%%%%%%%%%%%%%%%%%%

\section{Another effective model}
\label{AnotherModel}

Our general approach to constructing MHV rules can be tested in
a second interesting model with the effective operator,
\be
O_\Lambda = \frac{1}{\Lambda^2} 
\tr G_\mu^{~\nu} G_\nu^{~\rho} G_\rho^{~\mu} \,.
\label{GGGop}
\ee
This operator is the unique gauge-invariant, CP-even, dimension-6 
operator built solely from gluon fields (after applying equations of
motion).  Hence it provides a sensible way to characterize
possible deviations of gluon self-interactions from those predicted
by QCD, such as might be produced by gluon compositeness~\cite{Simmons}.
The operator $O_\Lambda$ also may be produced by integrating out
a heavy colored fermion (or scalar), albeit with a phenomenologically
tiny coefficient.
Various ways to probe for such an operator experimentally have been
proposed~\cite{Simmons,SCDDG,DS}.

Here we will consider the deviations that are linear in $1/\Lambda^2$;
that is, the set of (color-ordered) amplitudes $A_n^{(\Lambda)}$
arising from one insertion of $O_\Lambda$, combined with any number
of tree-level QCD interactions.  Curiously, at the four-parton level,
the amplitudes generated in this way are orthogonal to those of QCD.
For example, the four-gluon helicity amplitudes which are produced
are those which vanish in tree-level QCD, 
$({+}{+}{+}{+})$, $({-}{+}{+}{+})$, 
and the parity conjugates $({+}{-}{-}{-})$ and $({-}{-}{-}{-})$. 
At the linearized level, $O_\Lambda$ does not produce $({-}{-}{+}{+})$.
Thus the interference with tree-level QCD vanishes at this order.
Similarly, the $q\bar{q}gg$ amplitudes with only one power of
$O_\Lambda$ produce only $({\mp}{\pm}{+}{+})$ and the parity conjugate
$({\pm}{\mp}{-}{-})$, which vanish in QCD.
The five-gluon amplitudes which do interfere with QCD, $({-}{-}{+}{+}{+})$
and its parity conjugates, were computed in ref.~\cite{DS}.

The first on-shell $n$-gluon amplitudes generated by \eqn{GGGop} are those
for four gluons, which read, dropping an overall 
factor of $(-12i\pi/\Lambda^2)$~\cite{DS}:
\bea
A_4^{(\Lambda)}(1^-,2^-,3^-,4^-)\ &=&\
     - {2s_{12}s_{23}s_{13}\over\spb1.2\spb2.3\spb3.4\spb4.1}\ ,
     \label{Lambda-1}\\
  A_4^{(\Lambda)}(1^+,2^-,3^-,4^-)
  \ &=&\ {{\spa2.3}^2{\spa3.4}^2{\spa4.2}^2
     \over \spa1.2\spa2.3\spa3.4\spa4.1}\ ,
     \label{Lambda-2}\\
  A_4^{(\Lambda)}(1^-,2^-,3^+,4^+)
  \ &=& \  0\ ,\\
  A_4^{(\Lambda)}(1^-,2^+,3^-,4^+)
  \ &=&\  0\ ,
\eea
plus the parity conjugate amplitudes, 
$A_4^{(\Lambda)}(1^+,2^+,3^+,4^+)$ and
$A_4^{(\Lambda)}(1^-,2^+,3^+,4^+)$.

To apply our MHV-rules method to this model, we first rewrite 
the effective operator~(\ref{GGGop}) as the sum of a holomorphic
(selfdual) and an anti-holomorphic (anti-selfdual) term,
\be
O_\Lambda = \frac{1}{\Lambda^2} 
\left(\tr G_{{\sst SD}\,\mu}^{\,\,\,\,\,\,\,\,\,\,\,\,\nu} \,
G_{{\sst SD}\,\nu}^{\,\,\,\,\,\,\,\,\,\,\,\,\rho} \,
G_{{\sst SD}\,\rho}^{\,\,\,\,\,\,\,\,\,\,\,\,\mu}
\, + \,
\tr G_{{\sst ASD}\,\mu}^{\,\,\,\,\,\,\,\,\,\,\,\,\,\,\nu} \,
G_{{\sst ASD}\,\nu}^{\,\,\,\,\,\,\,\,\,\,\,\,\,\,\rho} \,
G_{{\sst ASD}\,\rho}^{\,\,\,\,\,\,\,\,\,\,\,\,\,\,\mu}
\right) \ .
\label{GGGop2}
\ee
The holomorphic interaction, $(G_{\sst SD})^3$, generates amplitudes with
a minimum of 3 negative-helicity gluons, such as
\eqns{Lambda-1}{Lambda-2}, whereas the anti-holomorphic interaction,
$(G_{\sst ASD})^3$, generates amplitudes with a minimum of 3 
positive-helicity gluons, such as $({-}{+}{+}{+})$ and $({+}{+}{+}{+})$.
(We can prove that the $(G_{\sst SD})^3$ amplitudes with 0 or 1
negative-helicity gluons vanish, along the same lines as the recursive
vanishing proof in the Higgs case in Appendix~\ref{DirectDemoSection},
using the structure of the vertex and \eqns{JJvanish}{EpsJJJvanish}.  
We assume that the $(G_{\sst SD})^3$ amplitudes with 2 negative-helicity 
gluons also vanish.)
That is, $(G_{\sst SD})^3$ leads to amplitudes with $n_-\ge 3$ 
and $n_+\ge 0$, whereas $(G_{\sst ASD})^3$ induces amplitudes with 
$n_+\ge 3$ and $n_-\ge 0$, as plotted in~\fig{TreeG3MapFigure}.

%%%%%%%%%%%%%%%%%%%%
%FIGURE
%
\FIGURE[t]{
{\epsfxsize 4.8 truein \epsfbox{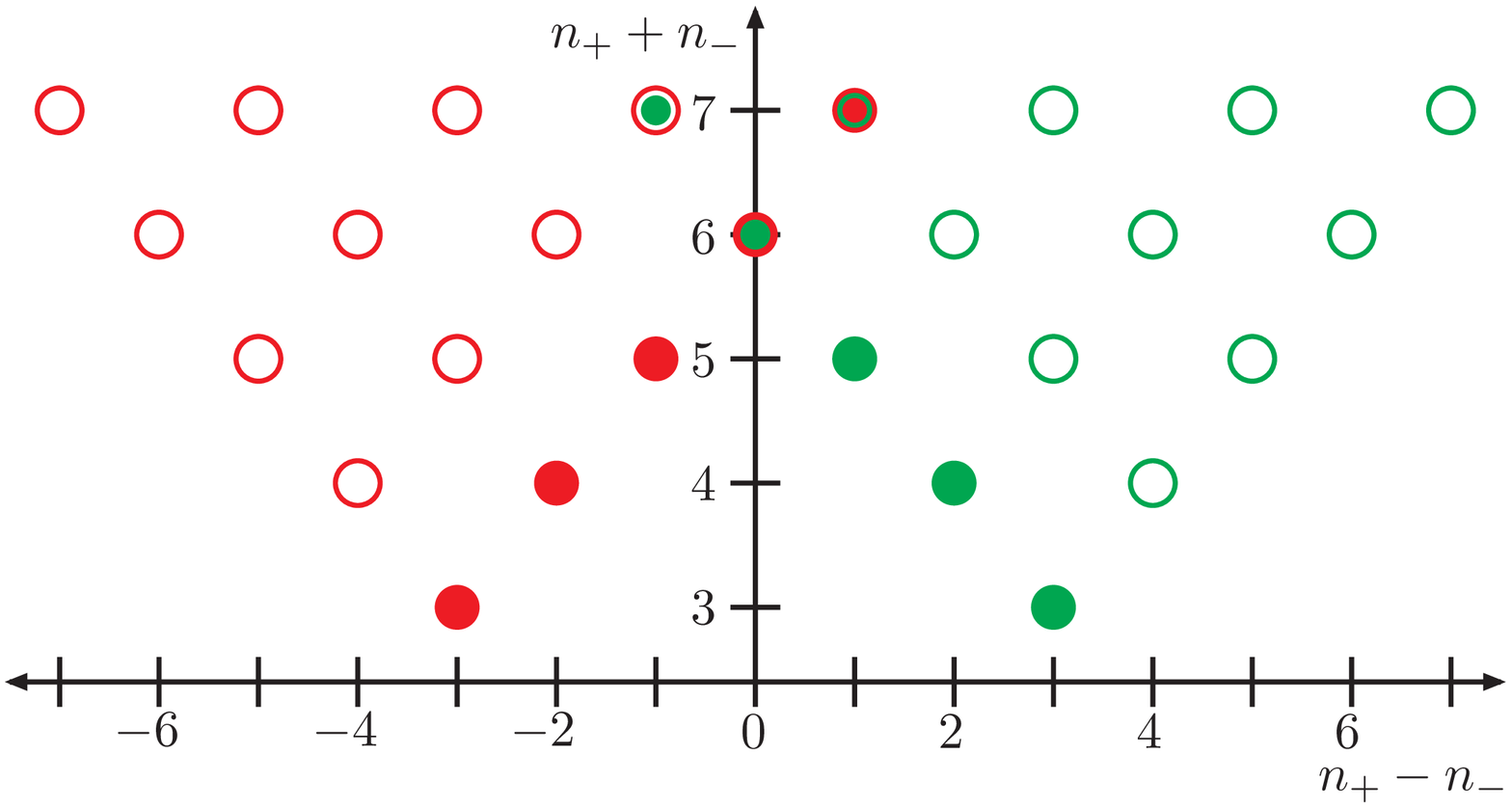}}
\caption{ The proposed structure of $n$-gluon amplitudes
induced by the operator $\tr(G^3)$.
The MHV tower for $\tr(G_{SD}^3)$-induced amplitudes contains pure MHV vertices 
(closed red circles) plus those amplitudes obtained by adding
pure-gauge-theory MHV vertices (open red circles).
The anti-MHV tower of $\tr(G_{ASD}^3)$-induced amplitudes
is obtained from the MHV tower by parity, and is shown in green.
The $\tr(G^3)$-induced amplitudes are given by the sum of the two
contributions, where they overlap.  The two entries with $n_+ + n_- = 3$
represent MHV vertices, but not on-shell scattering amplitudes.}
\label{TreeG3MapFigure}
}
%%%%%%%%%%%%%%%%%%%%%%

The building blocks induced by the holomorphic $(G_{\sst SD})^3$ 
interaction are the $G^3$-MHV vertices with $n_-= 3$ and arbitrary $n_+\ge 0$:
\be
A_n^{(\Lambda)}(1^+,\ldots,i^-,\ldots,j^-,\ldots,k^-,\ldots,n^+)
= { ( \spa{i}.{j} \spa{j}.{k} \spa{k}.{i} )^2
  \over \spa1.2 \spa2.3 \cdots \spa{n}.1 } \,,
\label{LambdaThreeMinus}
\ee
where only gluons $i,j,k$ have negative helicity.
On-shell, these are known to be correct for $n=4$ and 5~\cite{DS},
and they have the right factorization properties for $n>5$
(assuming the vanishing of $(G_{\sst SD})^3$ amplitudes with 
2 negative-helicity gluons).   We continue the spinor inner products 
off shell in the by now familiar way~\cite{CSW1}.
The anti-holomorphic $(G_{\sst ASD})^3$ interaction gives the 
$G^3$-anti-MHV vertices with $n_+= 3$ and arbitrary $n_-\ge 0$, 
which are the parity conjugates of \eqn{LambdaThreeMinus}.

The construction of the amplitudes induced by $\tr G^3$ closely
parallels the Higgs case.  To build the holomorphic MHV tower we use 
the $G^3$-MHV vertices~(\ref{LambdaThreeMinus}), combined with 
standard tree-level MHV vertices.  The anti-holomorphic tower
is its parity conjugate, built from $G^3$-anti-MHV vertices
combined with pure gauge theory anti-MHV vertices.  
\Fig{TreeG3MapFigure} depicts the two
towers.  In the overlap region (see \fig{TreeG3MapFigure})
we add the two terms.  In the $G^3$ case the overlap does not start 
until the 6-gluon amplitudes, $({-}{-}{-}{+}{+}{+})$, plus permutations.  

Just as in the Higgs case, this construction has the correct general
behavior under collinear and multi-particle factorization.  
Note that the $G^3$-MHV vertex~(\ref{LambdaThreeMinus}),
like the standard MHV vertex~(\ref{MHV}), has no ${-}{-} \to {-}$
or ${+}{-} \to {+}$ collinear singularity for $n>3$.
But it also is completely nonsingular in the collinear limit 
for $n=3$, so all the collinear singularities are described by the QCD
splitting amplitudes, as required~\cite{DS}.
Other than this consistency test, we have not carried out as extensive
checks of the $\tr G^3$ model as we did for the Higgs case, mainly
because only a limited number of amplitudes has been computed previously.
However, we have tested that the off-shell continuation of the 
3-point version of \eqn{LambdaThreeMinus}, namely 
$\spa{1}.{2} \spa{2}.{3} \spa{3}.{1}$, combined with the usual 
pure-gauge MHV $({-}{-}{+})$ vertex, successfully reproduces the
 $({-}{-}{-}{-})$ amplitude given in \eqn{Lambda-1}.

We can think of the effective interaction~(\ref{GGGop}) as 
being generated by integrating out a massive fermion or scalar
loop in a non-supersymmetric theory.  However,
it is still possible to embed the holomorphic plus anti-holomorphic 
decomposition~(\ref{GGGop2}) into an ${\cal N}=1$ supersymmetric
interaction,
\be
 {\cal L}^{\rm int} = \frac{1}{\Lambda^2}\,
 \int d^2\theta\  \tr [ (D^\beta\,W^\alpha)\,W_\beta\, W_\alpha ]~+~
{\rm h.\ c.}
\label{susyembd2}
\ee
The holomorphic part becomes a superpotential, exactly as in the Higgs model.
Hence we again have a supersymmetric completion  of an effective interaction,
which is an $F$-term.\footnote{%
Because the interaction is an $F$-term, it should not be
generated perturbatively in a supersymmetric theory.  This is why 
we took the microscopic theory, with a heavy fermion or scalar 
in the loop, to be nonsupersymmetric.}
However, in this case we have not been able to use
supersymmetric Ward identities to show vanishings of tree-level amplitudes with
less than 2 negative (or less than 2 positive) helicities. 
In fact, we know that such amplitudes, {\it e.g.} 
\eqns{Lambda-1}{Lambda-2}, do not vanish.  The reason why supersymmetric
Ward identities cannot be used in this model
is that the supersymmetry of the theory is spontaneously broken
in the presence of the interactions~(\ref{susyembd2}).  There is 
a term $D^3/\Lambda^2$ in \eqn{susyembd2} which, combined with the 
$D^2$ term from the SYM Lagrangian, leads the 
auxiliary $D$ field to develop a non-vanishing vacuum expectation value, 
$\langle D \rangle \propto \Lambda^2$.  Thus the 
supercharges do not annihilate the vacuum.

%%%%%%%%%%%%%%%%%%%%%%%%%%%%%%%%%%%

\section{Conclusions and outlook}
\label{ConclusionSection}

In this paper we have constructed and tested a novel set of MHV rules for
calculating scattering amplitudes of the massive Higgs boson plus an arbitrary
number of gluons. The model which we use to calculate these amplitudes is
the tree-level pure gauge theory plus an effective interaction
$HG_{\mu\nu}G^{\mu\nu}$.  This effective interaction is generated in the
heavy top quark limit from the non-supersymmetric Standard Model by
integrating out the heavy top-quark loop.  
To be able to apply MHV rules, we split the interaction into 
selfdual and anti-selfdual pieces.  The MHV rules lead
to compact formulae for the Higgs plus multi-parton amplitudes 
induced at leading order in QCD in the large $m_t$ limit.
We presented explicit formulae for the $\phi$-plus-$n$-gluon amplitudes 
containing up to four negative-helicity gluons, and an arbitrary number of
positive-helicity gluons.

This structure may also be useful for going to the next order in QCD ---
one-loop amplitudes for the Higgs boson plus many partons (or two loops if
we count the top-quark loop).  At present, such amplitudes are known
for up to three partons, namely the processes $Hggg$ and 
$Hgq\bar{q}$~\cite{Schmidt}, but the four-parton cases are 
required for the NLO weak-boson-fusion background computation
mentioned in the introduction.  One can split the computation
into $\phi$ and $\phi^\dagger$ terms.  Some of the helicity amplitudes 
should then become quite simple --- namely, those
for which the corresponding tree amplitudes vanish,
$A_n(\phi,1^\pm,2^+,\ldots,n^+)$.  These amplitudes must be rational
functions, free of all cuts, by the same argument as in the pure-gauge-theory
case~\cite{AllPlus,MahlonOneMinus}.  It may well be possible to determine them
for all $n$ in a similar fashion, by using recursive or 
collinear-based arguments.

Apart from being interesting on its own right, as discussed in the
introduction, we believe that this model gives important insights into the
structure of the MHV rules in generic nonsupersymmetric theories at the 
loop level.  In fact, the two examples of effective theories we have 
considered suggest a useful generalization.

Consider certain classes of loop diagrams in non-supersymmetric gauge
theories, such that loops can be integrated out and represented by
higher-dimensional operators in the effective action. The key idea for
constructing MHV rules for this effective action, is to split the
higher-dimensional operators into holomorphic and anti-holomorphic terms,
such that the holomorphic terms can be embedded into a superpotential of a
supersymmetric theory.  In other words, the holomorphic interactions will
involve only chiral superfields $W_\alpha$ and $\Phi$ (if matter is
present). In components, this means separating selfdual and anti-selfdual
components of the field strength.  Then we build the MHV tower by
combining new MHV vertices from the holomorphic superpotential with the
standard tree-level MHV vertices; the anti-MHV tower is built by combining
anti-MHV vertices from the anti-holomorphic interactions with the standard
anti-MHV vertices.

The MHV-rules construction described in this paper was designed to address
effective interactions at tree level. From the perspective of the
microscopic theory, our approach enables us to address only massive loops
in a non-supersymmetric theory. One of the main points we want to make is
that these MHV rules amount to more than adding a new class of vertices to
the tree-level rules of ref.~\cite{CSW1}, in particular, the two towers of MHV
and anti-MHV diagrams are crucial for the construction to work.  
In the soft-Higgs limit, $k_H\to0$, each tower becomes proportional
to the pure-gauge-theory tower, but the constant of
proportionality depends on the helicity content of the amplitude.
We expect that our findings will be useful in constructing MHV rules 
also for massless loops in nonsupersymmetric theories.

%%%%%%%%%%%%%%%%%%%%%%%%%%%%%%%%%%%

\acknowledgments We thank the authors of ref.~\cite{DFM} for providing 
us with a numerical program computing the amplitudes in their paper.
EWNG and VVK acknowledge PPARC Senior Fellowships.

\vskip0.4in
%\newpage

\appendix

%%%%%%%%%%%%%%%%%%%%%%%%%%%%%%%%%%%

\section{Conventions}
\label{Nots}

\subsection{Color}

The tree-level Higgs-plus-gluons amplitudes can be decomposed into
color-ordered partial amplitudes~\cite{DawsonKauffman,DFM} as
\begin{equation}
{\cal A}_n(H,\{k_i,\lambda_i,a_i\}) = 
i C g^{n-2}
\sum_{\sigma \in S_n/Z_n} \Tr(T^{a_{\sigma(1)}}\cdots T^{a_{\sigma(n)}})\,
A_n(H,\sigma(1^{\lambda_1},\ldots,n^{\lambda_n}))\,.
\label{TreeColorDecomposition}
\end{equation}
Here $S_n/Z_n$ is the group of non-cyclic permutations on $n$
symbols, and $j^{\lambda_j}$ labels the momentum $k_j$ and helicity
$\lambda_j$ of the $j^{\rm th}$ gluon, which carries the adjoint
representation index $a_i$.  The $T^{a_i}$ are fundamental
representation SU$(N_c)$ color matrices, normalized so that
$\Tr(T^a T^b) = \delta^{ab}$.  The strong coupling constant is
$\alpha_s=g^2/(4\pi)$.

Color-ordering means that, in a computation based on Feynman diagrams,
the partial amplitude
$A_n(H,1^{\lambda_1},2^{\lambda_2},\ldots,n^{\lambda_n})$
would receive contributions only from planar tree diagrams with a
specific cyclic ordering of the external gluons: $1,2,\ldots,n$.
Because the Higgs boson is uncolored, there is no color restriction on how 
it is emitted.  The partial amplitude $A_n$ is invariant under 
cyclic permutations of its gluonic arguments.
It also obeys a reflection identity, 
\be
A_n(H,n,n-1,\ldots,2,1) = (-1)^n A_n(H,1,2,\ldots,n-1,n),
\label{ReflectionIdentity}
\ee
and a dual Ward identity,
\be
A_n(H,1,2,3,\ldots,n) + A_n(H,2,1,3,\ldots,n)
+ \cdots + A_n(H,2,3,\ldots,n-1,1,n) = 0.
\label{DualWardIdentity}
\ee
These properties mimic those of the corresponding pure-gluon amplitudes
where the Higgs is omitted.

%%%%%%%%%%%%%%%%%%%%%%

\subsection{Spinors, helicity and selfduality}

We work in Minkowski space with the metric
$\eta^{\mu\nu}$  and use the sigma matrices from
Wess and Bagger~\cite{WessBagger},
$\sigma^\mu_{\alpha \dot\alpha}=(-1,\tau^1,\tau^2,\tau^3)$, and
$(\bar\sigma^{\mu})^{\dot\alpha \alpha}=(-1,-\tau^1,-\tau^2,-\tau^3)$, where
$\tau^{1,2,3}$ are the Pauli matrices.

In the spinor helicity formalism
\cite{SpinorHelicity} an on-shell momentum
of a massless particle, $k_\mu k^\mu=0$, is represented as
\be
k_{\alpha \dot\alpha} \equiv \ k_\mu \sigma^\mu_{\alpha \dot\alpha}
=\ \lambda_\alpha\tilde\lambda_{\dot\alpha} \ ,
\ee
where $\lambda_\alpha$ and $\tilde\lambda_{\dot\alpha}$
are two commuting spinors of positive and negative chirality.
Spinor inner products are defined
by\footnote{Our conventions for spinor helicities follow
refs.~\cite{Witten1,CSW1}, except that $[ij] = - [ij]_{CSW}$
as in ref.~\cite{LDTASI}.}
\be
\langle \lambda,\lambda'\rangle 
= \ \epsilon_{\alpha\beta}\lambda^\alpha\lambda'{}^\beta
 \,, \qquad
[\tilde\lambda,\tilde\lambda'] 
=\ -\epsilon_{\dot\alpha \dot\beta}
\tilde\lambda^{\dot\alpha}\tilde\lambda'{}^{\dot\beta} \,,
\label{Atwo}
\ee
and a scalar product of two null vectors,
$k_{\alpha \dot\alpha}=\lambda_\alpha \tilde\lambda_{\dot\alpha}$ and
$p_{\alpha \dot\alpha}=\lambda'_\alpha\tilde\lambda'_{\dot\alpha}$, becomes
\be \label{scprod}
k_\mu p^\mu =\ - {1\over 2}
\langle\lambda,\lambda'\rangle[\tilde\lambda,\tilde\lambda'] \,.
\ee
We use the shorthand $\spa{i}.{j}$ and $\spb{i}.{j}$ for the inner products of
the spinors corresponding to momenta $k_i$ and $k_j$,
\be
\spa{i}.{j} = \langle \lambda_i, \lambda_j \rangle \,,
\qquad
\spb{i}.{j} = [ \tilde\lambda_i, \tilde\lambda_j ].
\ee

For gluon polarization vectors we use
\be
\pol_\mu^\pm(k,\xi) = \pm { \langle \xi^\mp | \gamma_\mu | k^\mp \rangle
                 \over \sqrt{2} \langle \xi^\mp | k^\pm \rangle } \,,
\label{HelPol}
\ee
where $k$ is the gluon momentum and $\xi$ is the reference momentum, an
arbitrary null vector which can be represented as the product of two
reference spinors, 
$\xi_{\alpha\dot\alpha}=\xi_{\alpha}\tilde\xi_{\dot\alpha}$. 
We choose the reference momenta for all gluons to be the same, unless 
otherwise specified.  In terms of helicity spinors, 
$\pol_{\alpha \dot\alpha} = \pol^\mu (\gamma_\mu)_{\alpha \dot\alpha}$,
\eqn{HelPol} takes the form \cite{Witten1},
\bea
\pol_{\alpha \dot\alpha}^+ &=&
\sqrt{2} \frac{\xi_\alpha \tilde\lambda_{\dot\alpha}}
{\vev{\xi~\lambda}} \,,
\label{Pluspolvhel} \\ 
\pol_{\alpha \dot\alpha}^{-} &=& 
\sqrt{2} \frac{\lambda_\alpha \tilde{\xi}_{\dot\alpha}}
{[\tilde\lambda ~ \tilde{\xi}]} \,.
\label{Minuspolvhel}
\eea
To simplify the notation, we will drop the tilde-sign over the dotted 
reference spinor, so that 
$\xi_{\alpha \dot\alpha} = \xi_{\alpha}\xi_{\dot\alpha}$.

It follows from \eqn{HelPol}, or \eqns{Pluspolvhel}{Minuspolvhel}, that
\be
 \pol_i^+ \cdot \pol_j^+ = 0, \qquad \pol_i^- \cdot \pol_j^- = 0.
\label{PolDotVanish}
\ee

We define the dual field strength in Minkowski
space via
\be
{}^*G^{\mu\nu} = \ihf \epsilon^{\mu\nu\rho\sigma} G_{\rho\sigma} \ ,
\ee
and $\epsilon^{0123}=1=- \epsilon_{0123}$. The selfdual (SD) part of the field strength
is selected via $(\sigma^{\mu\nu})_{\alpha}^{\,\beta} G_{\mu\nu}$, where
\be
\sigma^{\mu\nu}= 
\frac{1}{4}(\sigma^{\mu}\bar\sigma^{\nu}-\sigma^{\nu}\bar\sigma^{\mu}) \ ,
\qquad
\sigma^{\mu\nu}=\frac{i}{2} \epsilon^{\mu\nu\rho\sigma} \sigma_{\rho \sigma} \ .
\ee
Similarly, the combination 
$(\bar\sigma^{\mu\nu})^{\dot\alpha}_{\,\dot\beta} G_{\mu\nu}$
gives rise to the anti-selfdual (ASD) part of the field strength. Here
\be
\bar\sigma^{\mu\nu}= 
\frac{1}{4}(\bar\sigma^{\mu}\sigma^{\nu}-\bar\sigma^{\nu}\sigma^{\mu}) \ ,
\qquad
\bar\sigma^{\mu\nu}=
-\frac{i}{2} \epsilon^{\mu\nu\rho\sigma} \bar\sigma_{\rho \sigma} \ .
\ee
In general, $G^{\mu\nu}$ can be written as a selfdual plus an 
anti-selfdual contribution,
\be
G^{\mu\nu}= \, (\sigma^{\mu\nu})_{\alpha}^{\,\beta}\, g^{\alpha}_{\,\beta} 
\, +\,
(\bar\sigma^{\mu\nu})^{\dot\alpha}_{\,\dot\beta}
\, {\tilde{g}}_{\dot\alpha}^{\,\dot\beta}
\ .
\ee
It is convenient to re-express the field strength in terms of spinor indices as
$G_{\alpha\dot\alpha \beta\dot\beta}=
\sigma^\mu_{\alpha \dot\alpha}\sigma^\nu_{\beta \dot\beta}
G_{\mu\nu}$. Then
the decomposition above reads as follows\footnote{%
In deriving \eqn{sddecomp} we have used the $\sigma$-matrix identities,
$\sigma^\mu_{\alpha \dot\alpha} \sigma^\nu_{\beta \dot\beta}
(\sigma^{\mu\nu})_{\gamma}^{\,\delta}
= \epsilon_{\dot\beta \dot\alpha}
 (\epsilon_{\gamma \alpha}\, \delta^{\delta}_{\,\beta}
 +\epsilon_{\gamma \beta}\, \delta^{\delta}_{\,\alpha})$, and
$\sigma^\mu_{\alpha \dot\alpha}\sigma^\nu_{\beta \dot\beta}
  (\bar\sigma^{\mu\nu})^{\dot\gamma}_{\,\dot\delta}
 = \epsilon_{\beta \alpha}
  (\epsilon_{\dot\beta \dot\delta}\, \delta^{\dot\gamma}_{\,\dot\alpha}
  +\epsilon_{\dot\alpha \dot\delta}\, \delta^{\dot\gamma}_{\,\dot\beta})$,},
\be
G_{\alpha\dot\alpha \beta\dot\beta} = \, \epsilon_{\dot\alpha \dot\beta}
\,(g_{\alpha\beta}+g_{\beta\alpha})-\epsilon_{\alpha \beta}
\,({\tilde{g}}_{\dot\alpha \dot\beta}+{\tilde{g}}_{\dot\beta \dot\alpha})
\ . \label{sddecomp}
\ee
{}From this one concludes that $\epsilon_{\dot\alpha \dot\beta}$ multiplies the
SD-component of the field strength, and $\epsilon_{\alpha \beta}$ multiplies
the ASD-component.
It then follows, as in ref.~\cite{Witten1},
that the ASD field strength corresponds to positive-helicity
gluons $g^{+}$ and the SD component gives negative-helicity gluons $g^{-}$.
To verify this, note that the linearized field strength
\be
G_{\alpha\dot\alpha \beta\dot\beta}=-i(k_{\alpha\dot\alpha} \pol_{\beta\dot\beta}
-k_{\beta\dot\beta} \pol_{\alpha\dot\alpha}) \,,
\label{AbelianFieldStrength}
\ee
evaluated for the positive polarization vector from \eqn{Pluspolvhel}, 
becomes proportional to 
$\tilde\lambda_{\dot\alpha}\tilde\lambda_{\dot\beta} 
(\lambda_\alpha \xi_\beta - \lambda_\beta \xi_\alpha )$.
Thus it contains only the $\epsilon_{\alpha \beta}$ (ASD) term.
Similarly, inserting the negative polarization vector~(\ref{Minuspolvhel})
into \eqn{AbelianFieldStrength} leads to a result containing only 
the $\epsilon_{\dot\alpha \dot\beta}$ (SD) term.

In a supersymmetric theory, the SD field strength 
$\sigma^{\mu\nu} G_{\mu\nu}$ enters the chiral superfield $W_\alpha$, 
and the ASD combination, $\bar\sigma^{\mu\nu} G_{\mu\nu}$, enters 
the anti-chiral superfield $\overline{W}^{\dot\alpha}$~\cite{WessBagger}.  
Since the SD field strength corresponds to a negative-helicity gluon, 
we will associate all component fields of {\it chiral} superfields 
with {\it negative} helicity particles, and those of {\it anti-chiral} 
superfields with {\it positive} ones.  Hence we have
\bea
&& W_\alpha = \{ g^{-},\lambda^{-} \} \ , \qquad \Phi 
= \{ \phi,\psi^{-} \} \ , \\
&& \overline{W}^{\dot\alpha} = \{g^{+}, \lambda^{+} \} \ , \qquad
\Phi^{\dagger} = \{\phi^{\dagger},\psi^{+}\} \ ,
\eea
where $g^{\pm}$ correspond to gluons with helicities $h=\pm 1$;
$\lambda^{\pm}$ are gluinos with $h=\pm 1/2$;
$\phi$ and $\phi^{\dagger}$ are complex scalar
fields; and $\psi^{\pm}$ are their fermionic superpartners.
If the mass $m_H$ of the chiral superfield vanishes, then
$\psi^{\pm}$ are $h=\pm 1/2$ helicity eigenstates.
For nonzero $m_H$ they are chirality, but not helicity, eigenstates.

%%%%%%%%%%%%%%%%%%%%%%%%%%%%%%%%

\section{Vanishing of $A_n(\phi,1^\pm,2^+,3^+,\ldots,n^+)$}
\label{VanishingSection}

In this appendix, we demonstrate that for the coupling
of $\phi$, there is nothing `more MHV' than the MHV amplitudes.
That is, we show that $A_n(\phi,g_1^\pm,g_2^+,g_3^+,\ldots,g_n^+)=0$.
We can do this in two ways, using supersymmetric Ward identities
and also more directly, via the Berends-Giele recursion
relations and off-shell currents~\cite{BerendsGiele}.
We also compute the non-vanishing amplitudes
$A_n(\phi^\dagger,1^+,2^+,3^+,\ldots,n^+)$ recursively.

%%%%%%%%%%%%%%%%

\subsection{Supersymmetry argument}
\label{SWIVanishingSection}

Since the $HGG$ interaction has a supersymmetric
completion~(\ref{susyembd}), we can use supersymmetric
Ward identities~\cite{SWI} to demonstrate vanishings of certain tree amplitudes.
Before proceeding, we write down the full supersymmetric Lagrangian,
\bea
{\cal L} &=& \int d^4\theta\ \Phi^\dagger \Phi
+ \int d^2\theta\ \biggl[ {m_H\over2} \Phi^2
          + {1\over4}  (1 - 4 C \Phi) \tr W^\alpha W_\alpha ]
\nonumber \\ 
&& \hskip2cm 
+ \int d^2\bar\theta\ \biggl[ {m_H\over2} \Phi^{\dagger2}
  + {1\over4} (1 - 4 C \Phi^\dagger) 
 \tr \overline{W}_{\dot\alpha}\overline{W}^{\dot\alpha} \biggr] 
\label{susyembdfull} \\
 &=& F^\dagger F - \del_\mu \phi^\dagger \del^\mu \phi 
      - i \bar\psi \delsl \psi 
+ {1\over2} \tr D^2 - {1\over4} \tr G_{\mu\nu} G^{\mu\nu}
- i \bar\lambda \Dsl \lambda
\nonumber \\ 
&& \hskip0.2cm 
 +\ \Biggl\{ m_H \biggl( F \phi - {1\over2} \psi^2 \biggr) 
- C \biggl[ - F \tr \lambda\lambda
  - 2 \sqrt{2} i \psi^\alpha 
 \tr \Bigl( \lambda_\alpha D  
       - (\sigma_{\mu\nu})_\alpha^{~\beta} \lambda_\beta 
               \, G^{\mu\nu}_{SD} \Bigr)
\nonumber \\ 
&& \hskip4.9cm 
  + \phi \tr \Bigl( - 2i \bar\lambda \Dsl \lambda 
                   - G_{{\sst SD}\,\mu\nu} G_{\sst SD}^{\mu\nu} \Bigr)
  \biggr]\ +\ \hbox{h. c.} \Biggr\}.
\label{susyembdfullcomp}
\eea
Because the term linear in the auxiliary field $D$ is also linear in 
the coefficient $C$, the $D$-term `potential' from integrating out $D$ 
is quadratic in $C$ and may be neglected.  On the other hand, the
$F$-term interaction has a linear term, 
\be
{\cal L}_F = - | m_H\phi - C \lambda\lambda |^2
  = - m_H^2 \phi^\dagger\phi 
    + C m_H ( \phi^\dagger \lambda\lambda
                     + \phi \bar\lambda\bar\lambda ) + \Ord(C^2).
\label{LFTerms}
\ee

To derive the supersymmetry Ward identities we will use the supersymmetry 
transformations of the on-shell fields in the helicity basis, 
following ref.~\cite{LDTASI},
\bea
\label{susyward}
&&[Q(\xi) \, , \, \lambda^{+}(k)] \ = \ - \theta \vev{\xi~k}\,g^+ (k) \ , 
\quad
[Q(\xi) \, , \, \lambda^{-}(k)] \ = \ + \theta [\xi~k]\,g^- (k) \ , \\
&&[Q(\xi) \, , \, g^{-}(k)] \ = \ + \theta \vev{\xi~k}\,\lambda^- (k) \ ,
\quad
[Q(\xi) \, , \, g^{+}(k)] \ = \ - \theta [\xi~k]\,\lambda^+ (k) \  , \\
&&[Q(\xi) \, , \, \phi^{\dagger}(k)] \ = \ - \theta \vev{\xi~k}\,\psi^+ (k) \ , 
\quad
[Q(\xi) \, , \, \phi(k)] \ = \ + \theta [\xi~k]\,\psi^- (k) \ , \label{Bsix} \\
&&[Q(\xi) \, , \, \psi^{-}(k)] \ = \ + \theta \vev{\xi~k}\,\phi (k) \ ,
\quad
[Q(\xi) \, , \, \psi^{+}(k)] \ 
= \ - \theta [\xi~k]\,\phi^{\dagger} (k) \ . \label{Bseven}
\eea
Here the operator $Q(\xi)$ is a Lorentz singlet entering a commutative
(rather than anticommutative) algebra
with all the fields. It is obtained from the standard spinor supercharge by
contracting it with a commuting reference spinor $\xi$ and
multiplying it by a Grassmann number $\theta$. This defines a
commuting singlet operator $Q(\xi)$.
In what follows, the anticommuting parameter
$\theta$ will cancel from the relevant expressions for the amplitudes.
The reference spinors, $\xi_\alpha$ and $\xi_{\dot \alpha}$,
are arbitrary spinors; they will be fixed below.

The supersymmetry relations involving $\phi$ and $\psi$,
\eqns{Bsix}{Bseven}, have been written for the massless Higgs case, $m_H=0$.
In principle, one can extend these relations in such a way that
they can also be applied in the case of a massive Higgs boson.
However, it turns out that the resulting SWI are not as useful for $m_H\neq0$,
so in the end we will only consider the massless Higgs case.

In order to prove that $A_n(\phi,g_1^\pm,g_2^+,g_3^+,\ldots,g_n^+)=0$
we consider the following equation:
\be
\langle \, 0 | \,[Q(\xi)\, , \, \phi_{k}\, g_{k_{1}}^{\pm}\,
\lambda^+_{k_2}\, g_{k_3}^+\, \ldots\, g_{k_n}^+ ]\, | \, 0 \, \rangle
\ = \ 0
\,.
\label{sysyw1}
\ee
The right-hand side is zero because, in a theory with unbroken supersymmetry,
the supercharge $Q$ annihilates the vacuum.

First, we consider \eqn{sysyw1} for the negative helicity choice $g_{k_{1}}^{-}$.
We find
\bea
0\,=\,&&
[\xi~k] \langle 0 |
\psi_{k}^{-}\, g_{k_{1}}^{-}\, \lambda^+_{k_2}\, g_{k_3}^+\, \ldots |0\rangle
\,+\,
\vev{\xi~k_1} \langle 0 |
\phi_{k}\, \lambda_{k_{1}}^{-}\, \lambda^+_{k_2}\, g_{k_3}^+\, \ldots |0\rangle
\nonumber \\
&& \hskip-0.2cm 
- \vev{\xi~k_2} \langle 0 |
\phi_{k}\, g_{k_{1}}^{-} g^+_{k_2}\, g_{k_3}^+\, \ldots |0\rangle
\,+\,
[\xi~k_3] \langle 0 |
\phi_{k}\, g_{k_{1}}^{-} \lambda_{k_2}^+\, \lambda_{k_3}^+\, \ldots |0\rangle
\,+\, \ldots
\label{susyw11}
\eea
The third term on the right hand side of \eqn{susyw11} is the amplitude we
want to investigate.  The remaining terms in \eqn{susyw11} contain one
fermion-antifermion pair. We will now set the reference spinor $\xi$ to be
equal to $k_1$ in order to discard the second term in \eqn{susyw11}. 
For $m_H=0$, we can drop the $\lambda$-chirality-violating 
$F$-terms~(\ref{LFTerms}).  
Then chirality conservation implies that $\lambda^+$ can only be in the same
amplitude with a $\lambda^-$ (our conventions are that all
particles are incoming), which kills the last set of terms.
Furthermore, there are no Feynman diagrams which can connect the 
$\psi^-$ fermion to the $\lambda^+$ fermion, because 
the $\psi \sigma_{\mu\nu} \tr \lambda \, G^{\mu\nu}_{SD}$
interaction (plus hermitian conjugate) in \eqn{susyembdfullcomp} 
is of the type,
\be
{\cal L}^{\rm int} \ni \psi^- \lambda^- g^- + \psi^+ \lambda^+ g^+ \,.
\label{chirviol}
\ee
Thus the first term also vanishes.  Because all the fermion-containing 
terms in~\eqn{susyw11} vanish, we conclude that the amplitude 
$A_n(\phi,g_1^-,g_2^+,g_3^+,\ldots,g_n^+)=0$.

We can similarly demonstrate that $A_n(\phi,g_1^+,g_2^+,g_3^+,\ldots,g_n^+)=0$
by starting with \eqn{sysyw1} and the positive-helicity gluon $g_{k_{1}}^{+}$.

At the same time, it is instructive to show that
when $\phi$ is exchanged with $\phi^\dagger$
the amplitudes with less than two negative-helicity gluons
can be non-vanishing. Here we will concentrate on amplitudes with
one negative-helicity gluon (amplitudes with no negative-helicity gluons
are proportional to $m_H^4$ and vanish in the massless limit we 
consider at present).  We proceed by considering a Ward identity,
\be
\langle \, 0 \, | \,[Q\, , \, \phi^\dagger_{k}\, g_{k_{1}}^{-}\,
\lambda^+_{k_2}\, g_{k_3}^+\, \ldots\, g_{k_n}^+ ]\, | \,  0\, \rangle
\ = \ 0 \,.
\label{sysyw2}
\ee
This gives us,
\bea
0\,=\,&&
-\vev{\xi~k} \langle 0 |
\psi_{k}^{+}\, g_{k_{1}}^{-}\, \lambda^+_{k_2}\, g_{k_3}^+\, \ldots |0\rangle
\,+\,
\vev{\xi~k_1} \langle 0 |
\phi^{\dagger}_{k}\, \lambda_{k_{1}}^{-}\, \lambda^+_{k_2}\, 
                                                 g_{k_3}^+\, \ldots |0\rangle
\nonumber \\
&& \hskip-0.0cm 
-\vev{\xi~k_2} \langle 0 |
\phi^{\dagger}_{k}\, g_{k_{1}}^{-} g^+_{k_2}\, g_{k_3}^+\, \ldots |0\rangle
\,+\,
[\xi~k_3] \langle 0 |
\phi^{\dagger}_{k}\, g_{k_{1}}^{-} \lambda_{k_2}^+\, 
                                         \lambda_{k_3}^+\, \ldots |0\rangle
\,+\, \ldots
\label{susyw22}
\eea
The third term on the right hand side of \eqn{susyw22} is the amplitude we
want to investigate.  The main difference with \eqn{susyw11} is that now
the first term on the right hand side of \eqn{susyw22} is non-vanishing,
since the $\psi^+ \lambda^+ g^+$ interaction is allowed due to
\eqn{chirviol}. This implies that 
$A_n(\phi^\dagger,g_1^+,g_2^+,g_3^+,\ldots,g_n^+) \neq 0$.

In the next part of this Appendix we recover these results without 
appealing to supersymmetric Ward identities or setting $m_H=0$.

%%%%%%%%%%%%%%%%

\subsection{Direct demonstration}
\label{DirectDemoSection}

We can also show that $A_n(\phi,1^\pm,2^+,3^+,\ldots,n^+)$ vanishes
using the Berends-Giele recursion relations and off-shell
currents~\cite{BerendsGiele}.  (For a review, see ref.~\cite{LDTASI}.)

The two-point vertex coupling $\phi$ to two (off-shell)
gluons with outgoing momenta $k_1$ and $k_2$ and Lorentz indices $\mu_1$
and $\mu_2$ is
\be
V^H_{\mu_1\mu_2}(k_1,k_2) =
\eta_{\mu_1\mu_2} k_1 \cdot k_2 - k_{1\,\mu_2} k_{2\,\mu_1}
 + i \pol_{\mu_1\mu_2\nu_1\nu_2} k_1^{\nu_1} k_2^{\nu_2} \,.
\label{PhiggVertex}
\ee
For $\phi^\dagger$ the sign of the Levi-Civita term would be reversed.

First let us compute the simplest amplitudes $A_2(\phi,1^\pm,2^\pm)$
using this vertex.
(The opposite-helicity cases vanish using angular-momentum conservation,
$A_2(\phi,1^\pm,2^\mp) = 0$.)
For gluon polarization vectors we use \eqn{HelPol}.
{}From identities \eqn{PolDotVanish} it follows that
only the second and third terms in the vertex~(\ref{PhiggVertex})
contribute to $A_2(\phi,1^\pm,2^\pm)$.  Consider the ratio of their
contributions in the positive-helicity case,
\bea
R^{++} &\equiv&
 { i \pol_{\mu_1\mu_2\nu_1\nu_2}
  \pol^{+,\mu_1} \pol^{+,\mu_2} k_1^{\nu_1} k_2^{\nu_2}
 \over - \pol_1^{+} \cdot k_2 \ \pol_2^{+} \cdot k_1 }
= - {1\over4} { \tr[ \gamma_5 \esl_1^+ \esl_2^+ \ksl_1 \ksl_2 ]
   \over \pol_1^{+} \cdot k_2 \ \pol_2^{+} \cdot k_1 }
\label{Rppdef} \\
&=& { \Bigl\{ \tr \Bigl[ \textstyle{{1\over2}}(1-\gamma_5)
  \gamma_{\mu_1} \gamma_{\mu_2} \ksl_1 \ksl_2 \Bigr]
 - \tr \Bigl[ \textstyle{{1\over2}}(1+\gamma_5)
  \gamma_{\mu_1} \gamma_{\mu_2} \ksl_1 \ksl_2 \Bigr] \Bigr\}
  \langle \xi^- | \gamma^{\mu_1} | k_1^- \rangle
  \langle \xi^- | \gamma^{\mu_2} | k_2^- \rangle
 \over 4 \, \spa{\xi}.2 \spb2.1 \, \spa{\xi}.1 \spb1.2 } \,.
\nonumber \\
&&~~~ \label{Phipp1}
\eea

Fierzing the $\langle \xi^- | \gamma^{\mu_i} | k_i^- \rangle$
strings into the trace gives
\bea
R^{++} &=&
 { \spa{\xi}.{\xi} \spb2.1 \spa1.2 \spb2.1
         - \spb1.2 \spa{\xi}.1 \spb1.2 \spa2.{\xi}
 \over \spa{\xi}.2 \spb2.1 \, \spa{\xi}.1 \spb1.2 }
\nonumber \\
&=&  -1.
\label{Phipp2}
\eea
Repeating the analysis for two negative-helicity gluons yields
the opposite sign,
\bea
R^{--} &\equiv&
 { i \pol_{\mu_1\mu_2\nu_1\nu_2}
  \pol^{-,\mu_1} \pol^{-,\mu_2} k_1^{\nu_1} k_2^{\nu_2}
 \over - \pol_1^{-} \cdot k_2 \ \pol_2^{-} \cdot k_1 }
\nonumber \\
&=& + 1.
\label{Phimm2}
\eea
Thus the second and third terms cancel in the positive-helicity case, so
$A_2(\phi,1^+,2^+) = 0$; whereas they add in the negative-helicity case,
for which one finds~\eqn{Hmm}.

To compute $A_n(\phi,1^\pm,2^+,3^+,\ldots,n^+)$ using
the Berends-Giele off-shell currents,
we merely join each gluon produced by the vertices from
$\phi \tr (G^{\mu\nu} - \tilde{G}^{\mu\nu})^2$
to an off-shell current.
The two currents we need are~\cite{BerendsGiele,LDTASI}
\be
J^{+,\mu}_{1,n} \equiv J^\mu(1^+,2^+,\ldots,n^+) =
{ \langle \xi^- | \gamma^\mu \Psl_{1,n} | \xi^+ \rangle
 \over \sqrt2 \spa{\xi}.1 \spa1.2 \cdots \spa{n-1,}.{n} \spa{n}.{\xi} }
\,,
\label{JAllPlus}
\ee
where all reference momenta are taken to be equal to $\xi$,
and
\be
J^{-,\mu}_{1,n} \equiv J^\mu(1^-,2^+,\ldots,n^+) =
{ \langle 1^- | \gamma^\mu \Psl_{1,n} | 1^+ \rangle
 \over \sqrt2 \spa1.2 \cdots \spa{n}.{1} }
 \sum_{m=3}^n { \langle 1^- | \ksl_m \Psl_{1,m} | 1^+ \rangle
                 \over P^2_{1,m-1} P^2_{1,m} }
\,,
\label{JOneMinus}
\ee
where the reference momentum choice is
$\xi_1 = k_2$, $\xi_2 = \cdots = \xi_n = k_1$.
In these formulae, $\Psl_{p,q} = k_p + k_{p+1} + \ldots + k_{q-1} + k_q$.

Actually, the current~(\ref{JOneMinus}) is not quite sufficient
for the proof in the one-minus case.  We really need the current
where the negative-helicity gluon appears at an arbitrary position
in the chain of positive-helicity gluons (all with the same reference
momentum), $J^\mu(2^+,3^+,\ldots,1^-,\ldots,n^+)$.  This current
has been constructed by Mahlon~\cite{MahlonOneMinus}.  The expression
is rather complicated, so we do not present it here.  It is sufficient
for our purposes to note that it is also proportional to 
$\la 1^- | \gamma^\mu \Psl_{1,n} | 1^+ \ra$.

For $A_n(\phi,1^+,2^+,3^+,\ldots,n^+)$, we take all reference momenta
equal to $\xi$, a generic vector.  For
$A_n(\phi,1^-,2^+,3^+,\ldots,n^+)$, we take $\xi_1 = k_2$,
$\xi_2 = \cdots = \xi_n = k_1 \equiv \xi$.  Then in both cases 
all the currents attaching
to the Higgs vertex are proportional to $\langle \xi^- | \gamma^\mu \ldots$.
In terms of spinor notation, all currents are proportional to $\xi_\alpha$.
This property is all we need to demonstrate (again via Fierz identities) that
\bea
  J^+ \cdot J^+ = J^+ \cdot J^- &=& 0,
\label{JJvanish} \\
   \pol_{\mu_1\mu_2\mu_3\mu_4} J^{+,\mu_1} J^{+,\mu_2} J^{-,\mu_3}
 &=& 0.
\label{EpsJJJvanish}
\eea
These relations in turn suffice to show that the
Feynman vertices coupling $\phi$ to 3 or 4 gluons, $\phi ggg$ and $\phi gggg$,
do not contribute to $A_n(\phi,1^\pm,2^+,3^+,\ldots,n^+)$.
Terms in these vertices without a Levi-Civita tensor always attach
a Minkowski metric $\eta_{\mu_1\mu_2}$ to two currents; their
contribution vanishes according to \eqn{JJvanish}.
(The same is true of the first term in the $\phi gg$
vertex~(\ref{PhiggVertex}).)
Terms containing the Levi-Civita tensor $\pol_{\mu_1\mu_2\mu_3\mu_4}$
attach it directly to at least three currents;
their contribution vanishes according to \eqn{EpsJJJvanish}.
This leaves just the contributions of the second and third terms
in the $\phi gg$ vertex~(\ref{PhiggVertex}).
They cancel against each other, just as in the case of
$A_2(\phi,1^+,2^+)$ above.  Suppose that gluons $p+1$ through $m$
(cyclicly) attach to one leg of the $\phi gg$ vertex, and gluons
$m+1$ through $p$ (cyclicly) attach to the other leg.
Then the ratio analogous to \eqn{Rppdef} is
\bea
R^{*,\pm +}_{p,m} &\equiv&  { i \pol_{\mu_1\mu_2\nu_1\nu_2}
  \langle \xi^- | \gamma^{\mu_1} \Psl_{p+1,m} | \xi^+ \rangle
  \langle \xi^- | \gamma^{\mu_2} \Psl_{m+1,p} | \xi^+ \rangle 
   k_1^{\nu_1} k_2^{\nu_2}
 \over - \langle \xi^- | \Psl_{m+1,p} \Psl_{p+1,m} | \xi^+ \rangle
         \langle \xi^- |  \Psl_{p+1,m} \Psl_{m+1,p} | \xi^+ \rangle }
\label{Rmndef} \\
&=& - 1,
\label{Rmnanswer}
\eea
using the same Fierz identities as before.
This completes the recursive proof that
\be
 A_n(\phi,1^\pm,2^+,3^+,\ldots,n^+) = 0.
\label{SUSYRecursivevanish}
\ee

%%%%%%%%%%%%%%%%%%%%%

\subsection{Recursive construction of $A_n(\phi^\dagger,1^+,2^+,\ldots,n^+)$}
\label{RecursiveAllPlusSection}

Using the same all-plus current~(\ref{JAllPlus}), but 
flipping the sign of the Levi-Civita term in the Higgs 
vertex~(\ref{PhiggVertex}), we can give a recursive construction of 
the non-vanishing all-plus amplitudes 
$A_n(\phi^\dagger,1^+,2^+,\ldots,n^+)$.  Note that 
from the point of view of MHV rules, these amplitudes are 
`maximally googly'.

The only difference from the above analysis of
$A_n(\phi,1^+,2^+,3^+,\ldots,n^+)$ is that now the ratio
$R^{*,++}_{p,m}$ in \eqn{Rmnanswer} becomes equal to $+1$,
so the terms coming from the second and third terms in the 
vertex~(\ref{PhiggVertex}) add instead of cancelling.  
Restoring the overall factors,
\bea
A_n^{\dagger+\cdots+} \equiv
A_n(\phi^\dagger,1^+,2^+,\ldots,n^+) 
&=& - {1\over \rho} \sum_{1\leq m < p \leq n}
  { \spa{m,}.{m+1} \over \spa{m}.{\xi} \spa{\xi}.{m+1} }
  { \spa{p,}.{p+1} \over \spa{p}.{\xi} \spa{\xi}.{p+1} }
\nonumber \\
&& \hskip2.0cm
 \times \la \xi^- | \Psl_{p+1,m} \Psl_{m+1,p} | \xi^+ \ra^2 \,,
\label{NonvanishAllplus1}
\eea
where $\rho = \spa1.2 \spa2.3 \cdots \spa{n}.{1}$.
We can replace $\Psl_{p+1,m}$ by $\Psl_{1,n}$, and then
expand out $\Psl_{m+1,p}$ to get,
\bea
A_n^{\dagger+\cdots+} 
&=& - {1\over \rho} \sum_{1\leq m < (k,l) \leq p \leq n}
  { \spa{m,}.{m+1} \over \spa{m}.{\xi} \spa{\xi}.{m+1} }
  { \spa{p,}.{p+1} \over \spa{p}.{\xi} \spa{\xi}.{p+1} }
        \la \xi^- | \Psl_{1,n} \ksl | \xi^+ \ra
        \la \xi^- | \Psl_{1,n} \lsl | \xi^+ \ra \,.
\nonumber \\
&&~ \label{NonvanishAllplus2}
\eea
We use the $k \lr l$ symmetry to write the sum over $k$ and $l$
as twice the sum over $m < k < l \leq p$,
plus the (diagonal) sum over $m < k = l \leq p$.
We then apply the eikonal identity,
\be 
 \sum_{m=j}^{k-1} { \spa{m,}.{m+1} \over \spa{m}.{\xi} \spa{\xi,}.{m+1} }
  = { \spa{j}.{k} \over \spa{j}.{\xi} \spa{\xi}.{k} } \,,
\label{EikonalId}
\ee
in order to carry out the sums over $m$ and $p$.  After rearranging
some of the spinor products, we have
\bea
A_n^{\dagger+\cdots+} 
&=& {1\over \rho {\spa{\xi}.{1}}^2 } \biggl\{ 
 2 \sum_{1\leq k < l \leq n} 
   \la \xi^- | \Psl_{1,n} \ksl | 1^+ \ra
   \la \xi^- | \Psl_{1,n} \lsl | 1^+ \ra
+ \sum_{1\leq k \leq n} \la \xi^- | \Psl_{1,n} \ksl | 1^+ \ra^2 
  \biggr\}
\nonumber \\
&=&  {1\over \rho {\spa{\xi}.{1}}^2 } 
         \la \xi^- | \Psl_{1,n} \Psl_{2,n} | 1^+ \ra^2 
\nonumber \\
&=&  { P_{1,n}^2 \over \rho } 
\nonumber \\
&=&  { m_H^4 \over \spa1.2 \spa2.3 \cdots \spa{n}.{1} } \,,
\label{NonvanishAllplus3}
\eea
as desired.  Since the corresponding $\phi$ amplitude vanishes,
\eqn{NonvanishAllplus3} is also the result for the all-plus Higgs 
amplitude, \eqn{HAllPlus}.

%%%%%%%%%%%%%%%%%%%%%%%%%%%%%%%%%%%%%%%%%%%%%%%%

\end{document}